# Steady-state entanglement of spin qubits mediated by nonreciprocal and chiral magnons

Martijn Dols,[1,*] Mikhail Cherkasskii,[1] Victor A. S. V. Bittencourt,[2] Carlos Gonzalez-Ballestero,[3] Durga B. R. Dasari,[4] and Silvia Viola Kusminskiy[1,5,†]

[1] *Institute for Theoretical Solid State Physics, RWTH Aachen University, 52074 Aachen, Germany*
[2] *Institut de Science et d'Ingénierie Supramoléculaires (ISIS, UMR7006), Université de Strasbourg, 67000 Strasbourg, France*
[3] *Institute for Theoretical Physics and Vienna Center for Quantum Science and Technology, TU Wien, 1040 Vienna, Austria*
[4] *3. Physikalisches Institut, ZAQuant, University of Stuttgart, Allmandring 13, 70569 Stuttgart, Germany*
[5] *Max Planck Institute for the Science of Light, Staudtstraße 2, 91058 Erlangen, Germany*

We propose a hybrid quantum system in which a magnet supporting nonreciprocal magnons, chiral magnons, or both mediates the dissipative and unidirectional coupling of spin qubits. By driving the qubits, the steady state of this qubit-qubit coupling scheme becomes the maximally entangled Bell state. We devise a protocol where the system converges to this entangled state and benchmark it including qubit decay and dephasing. The protocol is numerically tested on a hybrid system consisting of nitrogen-vacancy (NV) centers coupled to magnon surface modes of a yttrium iron garnet (YIG) film. We show that the dephasing time of the NV centers forms the bottleneck for achieving the entanglement of NV centers separated by a distance within the magnon coherence length. Our findings identify the key technological requirements and demonstrate a viable route toward steady-state entanglement of solid-state spins over distances of several microns using magnonic quantum networks, expanding the toolbox of magnonics for quantum information purposes.

## I. INTRODUCTION

The capability of tailoring and controlling the interactions between quantum emitters (QEs) is an essential requisite for functional quantum networks. Such interactions can be engineered to yield unidirectional coupling, where two QEs couple asymmetrically to each other such that the dynamics of one QE influences the other, but not the other way around [1,2]. This type of coupling can be harnessed to design quantum networks in which information flows in only one direction, with advantages for applications in information processing [3,4]. In particular, it can be used for the generation of steady-state entanglement [5,6], which is not achievable for the bidirectional equivalent [7,8]. Achieving high-fidelity entanglement is a required resource for quantum information protocols such as cryptography [9], error correction [10,11], and teleportation [12]. (Uni)directional coupling can be realized via modes exhibiting direction-dependent polarization [3] combined with QEs exhibiting polarization-dependent transitions. Such setup has been realized, for example, in quantum optics, by coupling atoms and quantum dots to waveguides and cavities tailored to exhibit fields with direction-locked polarization [13–22].

The integration of optical systems typically constrains some of the tunability of the hybrid system [23]. A promising alternative seeks to incorporate magnons, i.e., the quanta of spin waves in magnetic systems, as information carriers exhibiting *in situ* tunability as well as scalability to nanoscale devices [24–29]. Magnons can be integrated seamlessly in hybrid systems since they can be coherently coupled to a vast range of excitations, including phonons [30–35], microwaves [36–40], optical photons [41–45], superconducting circuits [46–51], and single-spins systems [52–64]. Crucially to this work, magnons can show intrinsic nonreciprocity and chirality, which are two central ingredients to realize unidirectional coupling between QEs. Specifically, the magnetic potential of the spin waves can depend on the propagation direction. This has been demonstrated, e.g., for surface spin waves in ferromagnetic slabs [65,66], giving rise to the so-called field displacement nonreciprocity. Such surface waves are inherently also chiral; i.e., they exhibit polarization-direction locking. Spin waves can also exhibit a dispersion relation that is asymmetric with respect to the direction propagation, a phenomenon called frequency nonreciprocity. Systems possessing this trait include ferromagnetic ultrathin layers [67–71], ferromagnetic nanotubes [72,73], and ferromagnetic [74–79] and antiferromagnetic [80–84] multilayered structures. Using magnons as mediators for spin-spin coupling has been proposed to realize entanglement between spin qubits [52,54,57,59,60,64], including (uni)directional spin-spin coupling mediated by topological magnons [59] and chiral magnons [64]. However, the entanglement between spin qubits in these works is time dependent, as opposed to stationary.

*Contact author: martijn.dols@rwth-aachen.de
†Contact author: kusminskiy@physik.rwth-aachen.de

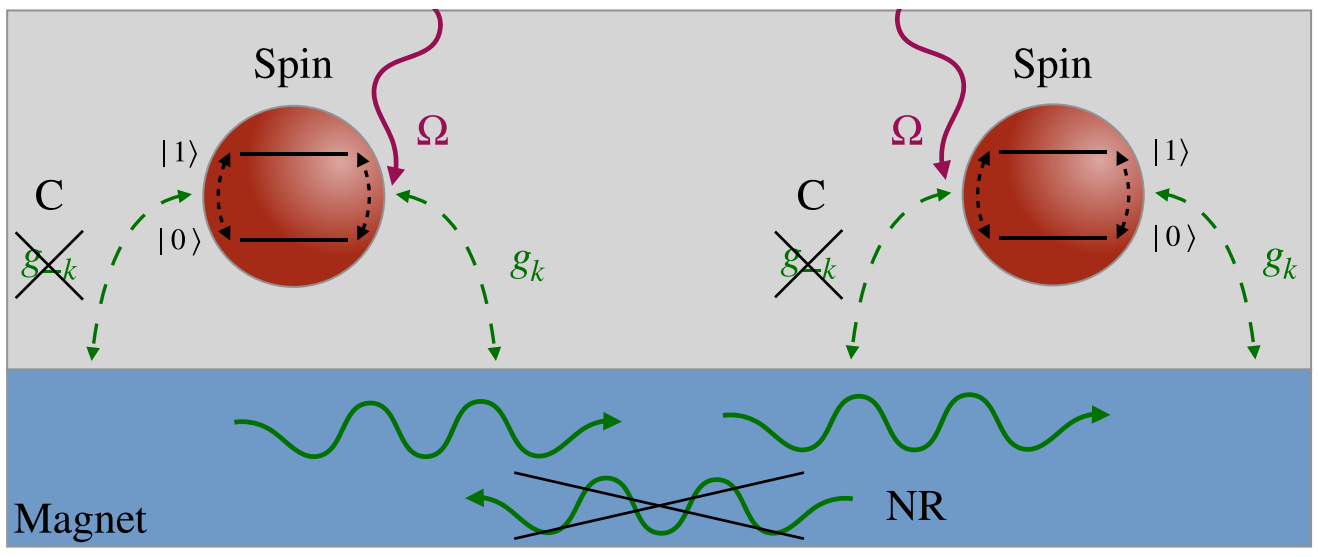

FIG. 1. Unidirectional coupling scheme between spin qubits mediated by magnons. The nonreciprocity (NR) of the magnet limits the propagation of the surface magnons to one propagation direction. Due to the chirality (C), the spin-magnon coupling $g_{\boldsymbol{k}}$ is nonvanishing only for one propagation direction of the magnons. The spin qubits are resonantly driven with a Rabi frequency Ω.

In this work, we demonstrate that the nonreciprocity and chirality of magnons can be exploited to generate steady-state entanglement between remote spin qubits. Specifically, we consider a hybrid system (Fig. 1) in which two spin qubits are coupled to the spin-wave modes of a magnetic material that supports nonreciprocal and chiral magnon propagation. Under continuous resonant driving of the qubits, the magnon-mediated interaction causes any initial two-qubit state to evolve toward a Bell state, i.e., a maximally entangled state. The steady-state character of the generated entanglement is advantageous for quantum networking applications, because the target state is a stationary attractor of the dynamics rather than a transient state.

Building on the theoretical result, we consider a specific protocol to prepare the target state. We optimize the protocol by minimizing the time required to reach a given fidelity threshold, i.e., the time to approach the entangled state up to a set fidelity, and investigate the impact of intrinsic qubit relaxation and dephasing. Our analysis benchmarks the threshold temperature as well as the minimal dephasing and decay times required to achieve high-fidelity entangled states, providing guidelines for experimental design.

We assess a practical implementation of the proposed magnon-mediated entanglement scheme by considering a state-of-the-art platform: Two nitrogen-vacancy (NV)-center qubits [85,86] coupled via the magnetostatic surface magnons of a yttrium iron garnet (YIG) film [53,55,61]. We find that high-fidelity entanglement can be achieved in this setup provided that (1) the separation of the NV centers is within the magnon coherence length, on the order of a few millimeters for YIG at low temperature [87], and (2) the device is operated at cryogenic temperature, required to suppress thermal magnon population. Under these conditions, the main limitation for steady-state entanglement becomes the dephasing time of the NV centers. We find that a minimal dephasing time on the order of $T_\phi \sim 1.5$ s is required. This is an order of magnitude longer than the typical dephasing time of NV centers in current experiments, identifying the need for improved coherence preservation (e.g., via dynamical decoupling techniques [88]). Nevertheless, NV coherence times approaching the few-second regime have been reported [89], and further enhancements would make the envisioned entanglement protocol feasible.

The remainder of this work is structured as follows. In the first part of the article, we introduce the general concepts and the entangling protocol. In the second part, we propose the physical implementation of the protocol using state-of-the-art magnetic platforms. Specifically, in Sec. II we show the derivation the unidirectional coupling of (spin) qubits mediated by a magnon bath. The proposed protocol and its benchmark of decoherence, as well as the benchmark of directional over unidirectional coupling, are introduced in Sec. III. We discuss the implementation of chiral magnons in Sec. IV A and of nonreciprocal magnons in Sec. IV B. In Sec. IV C, we consider a particular implementation consisting of NV centers and a YIG film. In Sec. V, we present the conclusions and discuss possible alternative platforms. We relegate long derivations and further characterization and benchmarking of the protocol to the Appendixes. We derive in detail the master equation describing the unidirectionally coupled qubits for zero and finite temperature in Appendixes A and B, respectively. We present the eigenstates of the effective Hamiltonian of this master equation in Appendix C. In Appendix D, we discuss how the protocol time compares to the transient time of the steady state and dissipation effects in Appendix G. We provide benchmarks of the inverse entangled-state fraction, the temperature, and the qubit-qubit distance in Appendixes E, F, and H, respectively. The magnetic stray field of the YIG film, which is used in Sec. IV C, is derived in Appendix I.

## II. MODEL

The low-energy excitations of magnetic systems are bosonic quasiparticles known as magnons. In the spin-wave limit without considering magnon-magnon interaction terms, the Hamiltonian is generally given by

$$\hat{H}_M = \sum_{\boldsymbol{k}} \hbar\omega_{\boldsymbol{k}} \hat{m}_{\boldsymbol{k}}^\dagger \hat{m}_{\boldsymbol{k}}, \tag{1}$$

where the operator $\hat{m}_{\boldsymbol{k}}^{(\dagger)}$ annihilates (creates) a magnon in mode $\boldsymbol{k}$ with frequency $\omega_{\boldsymbol{k}}$. The magnon modes and their dispersion depend on the geometry of the magnetic system considered.

We consider that magnons couple to identical qubits that are classically driven with a driving frequency $\omega_d$ and Rabi frequency Ω. The two states $|0\rangle$ and $|1\rangle$ of the qubits are separated by an energy $\hbar\omega_q$. In the rotating frame of the drive and within the rotating wave approximation (RWA), valid for $\Omega \ll \omega_q + \omega_d$, the Hamiltonian reads

$$\hat{H}_Q = \sum_i \hbar\delta_q \hat{\sigma}_i^+ \hat{\sigma}_i^- + \frac{\hbar\Omega}{2}(\hat{\sigma}_i^+ + \hat{\sigma}_i^-), \tag{2}$$

with the lowering operator $\hat{\sigma}^- = |0\rangle\langle 1|$, the raising operator $\hat{\sigma}^+ = (\hat{\sigma}^-)^\dagger$, and the detuning $\delta_q = \omega_q - \omega_d$. We note that the usage of $\hat{\sigma}_i^+\hat{\sigma}_i^-$ and the Pauli operator $\hat{\sigma}_i^z$ is equal to each other up to an irrelevant constant.

The magnetic-field fluctuations generated by the magnons interact with the magnetic moment of the spin qubits, giving rise to a dipolar coupling. Under the RWA, this resonant interaction is described by [52,54–64]

$$\hat{H}_{\text{int}} = \sum_{i,\boldsymbol{k}} \hbar (g_{\boldsymbol{k},i}\hat{\sigma}_i^+ \hat{m}_{\boldsymbol{k}} + g^*_{\boldsymbol{k},i}\hat{m}^\dagger_{\boldsymbol{k}}\hat{\sigma}_i^-). \tag{3}$$

The directionality of the coupling is embedded in the coupling constants $g_{\boldsymbol{k},i}$. In a magnetic system, unidirectionality can arise due to chiral or unidirectional magnon transport (see Sec. IV). The consequence of both scenarios is that the qubits couple to only one specific propagation direction of the magnons, as visualized in Fig. 1, where the qubits are positioned along the axis where the time-reversal symmetry is broken [90,91], which we refer to as the unidirectional axis.

We derive the effective qubits' dynamics by tracing out the magnons within the Born-Markov approximation in Appendix A. Assuming that the magnons remain at zero temperature, we obtain an effective master equation describing the evolution of the qubits density operator $\hat{\rho}$, i.e., $\frac{\partial \hat{\rho}}{\partial t} = \mathcal{L}[\hat{\rho}]$, with the Liouvillian

$$\mathcal{L}[\hat{\rho}] = -\frac{i}{\hbar}([\hat{H}_Q, \hat{\rho}] + \hat{H}_{\text{eff}}\hat{\rho} - \hat{\rho}\hat{H}^\dagger_{\text{eff}}) + \hat{L}\hat{\rho}\hat{L}^\dagger \tag{4}$$

and an effective non-Hermitian Hamiltonian $\hat{H}_{\text{eff}} = \hat{H}_{\text{loc}} + \hat{H}_{\text{uni}}$.

The first term of this effective Hamiltonian describes the dissipation of the qubit due to the magnon bath:

$$\hat{H}_{\text{loc}} = -i\hbar \sum_i \frac{J_q}{2}\hat{\sigma}_i^+\hat{\sigma}_i^-, \tag{5}$$

where $J_q = 2\pi\mathcal{D}(k_q)|g_{k_q}|^2/|v_{k_q}|$ is the magnon-mediated dissipative coupling constant. $\mathcal{D}(k)$ and $v_k = \partial\omega/\partial k$ are the magnonic density of states and the group velocity, respectively, evaluated at the wave number $k_q$ that is in resonance with the qubit, i.e., $\omega_{k_q} = \omega_q$. Here, we assumed $v_{k_q} > 0$ with the wave number defined along the unidirectional axis, where we set the notation $\boldsymbol{k} \to k$ for the one-dimensional case. This Hamiltonian acts locally on the qubits and hence does not lead to interqubit coupling.

The limitation of the coupling to only one propagation direction (chiral case), or the existence of only one propagation direction (nonreciprocal case), along the unidirectional axis generates a unidirectionality in the coupling between qubits, which is described by the second term of the effective Hamiltonian

$$\hat{H}_{\text{uni}} = -i\hbar \sum_{i>j} J_q e^{ik_q r_{i,j}}\hat{\sigma}_j^-\hat{\sigma}_i^+, \tag{6}$$

with the distance $r_{i,j} = r_i - r_j$ between qubits $i$ and $j$. The ordering $i > j$ of the qubits in Eq. (6) is dictated by the position along the unidirectional axis, such that $r_i > r_j$. For details on the derivation, see Eqs. (A7)–(A18) in Appendix A.

The last term of the Liouvillian in Eq. (4) corresponds to a quantum jump term with the Lindblad operator

$$\hat{L} = \sum_i \sqrt{J_q} e^{-ik_q r_i}\hat{\sigma}_i^-. \tag{7}$$

The lack of a Lindblad term describing the thermal excitation of the qubits is a consequence of the zero-temperature model. Taking a bath with a finite temperature such a term arises, as shown in Eq. (B10) in Appendix B.

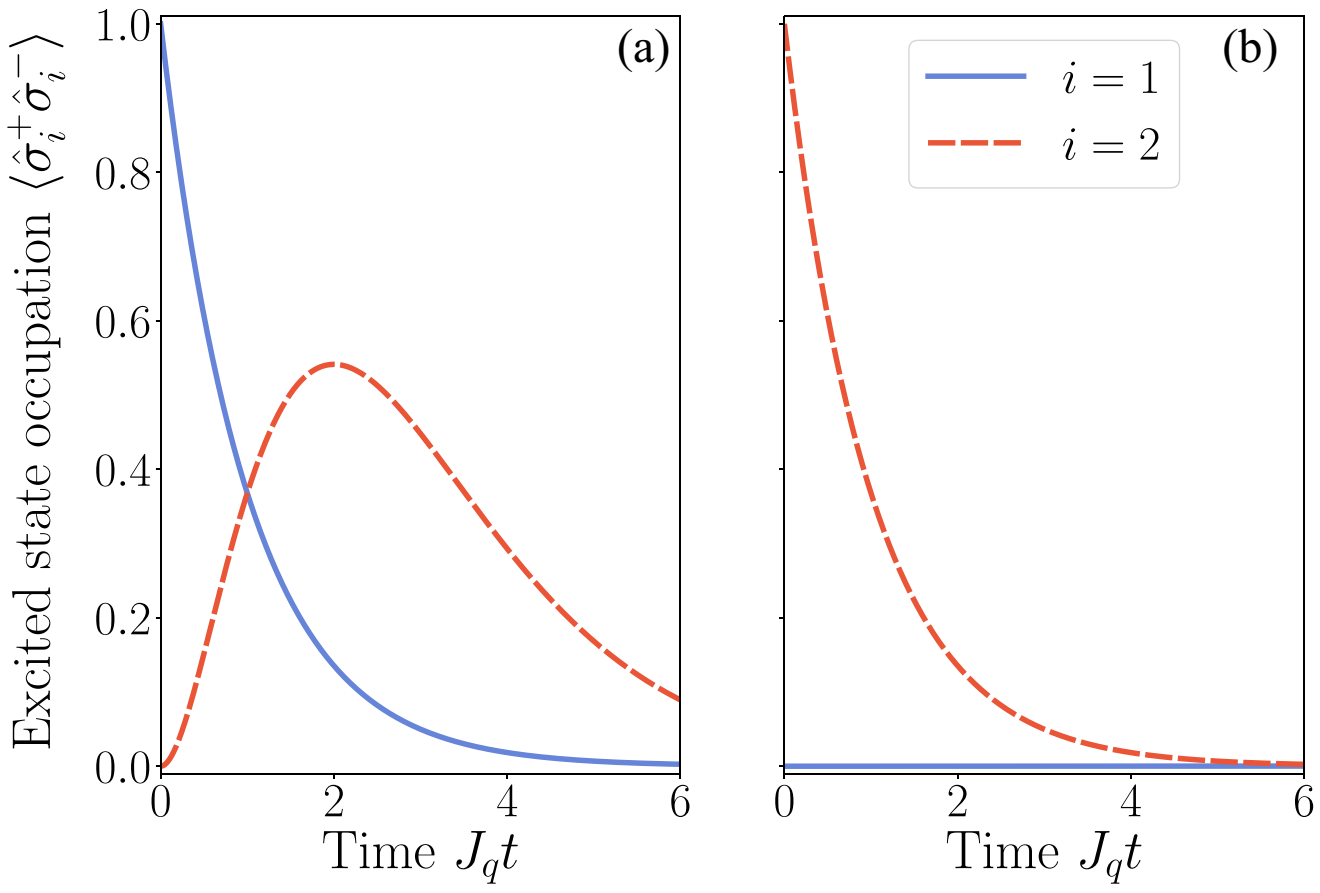

FIG. 2. Time dynamics according to the Liouvillian in Eq. (4) with initial state $|10\rangle$ in panel (a) and $|01\rangle$ in panel (b). The occupation of the excited state $\langle\hat{\sigma}_i^+\hat{\sigma}_i^-\rangle$ of qubit $i = 1, 2$ is shown as a function of the time $t$ normalized by the coupling $J_q$. For the input state $|10\rangle$ in panel (a), there is a transfer of excitation from qubit 1 to qubit 2, on top of the dissipation due to the bath. This transfer is lacking in panel (b), where the input state is $|01\rangle$, showing the unidirectionality of the coupling. In this figure, we set $J_q = 25\,\Omega$.

Equation (4) is valid provided the Markov approximation is fulfilled, which is ensured by the conditions $\tau_m \ll \tau_q,\ \tau_{\delta,\Omega}$, where $\tau_m$ and $\tau_q$ are the correlation times of the magnon bath and the spin qubit, respectively, and $\tau_{\delta,\Omega} = 2/\sqrt{\delta^2 + \Omega^2}$ is the time constant associated with the driving of the qubit. Also, we assumed that the coupling between qubits $i$ and $j$ is instantaneous following the Markov approximation, where the dynamics of the qubits is assumed to be slow with respect to the bath. Since the qubit dynamics are controlled by the frequencies $|\delta_q|$, $J_q$, and $\Omega$ as per the Liouvillian of Eq. (4), while the rate at which the qubits $i$ and $j$ exchange information through the bath is given by $v_{k_q}/r_{i,j}$, the instantaneous coupling condition reads $|\delta_q|,\ J_q,\ \Omega \ll v_{k_q}/r_{i,j}$. Finally, the qubit-qubit distance $r_{i,j}$ may not exceed the coherence length of the magnons, given by $l_m = v_{k_q}\tau_m$, corresponding to the condition $r_{i,j}/(v_{k_q}\tau_m) \ll 1$.

The unidirectionality in the coupling between the qubits is best reflected when considering the time dynamics of the two-qubit initial states $|10\rangle$ and $|01\rangle$. As prescribed by Eq. (6), an excitation is only transferred from qubit 1 to qubit 2, if qubit 2 is located after qubit 1 on the unidirectional axis. Therefore, we expect that there is a transfer of quanta only for the initial state $|10\rangle$ and not for $|01\rangle$. This expectation is confirmed in Fig. 2, where we plot the evolution in time of the excited state occupation according to Eq. (4). Due to the dissipative coupling given by Eq. (5), there will always be loss of the excitation of qubit $i = 1, 2$ on top of the unidirectional coupling.

We show the eigenstates of the Hamiltonian $\hat{H}_Q + \hat{H}_{\text{eff}}$ for two qubits driven resonantly in Appendix C. We demonstrate that one of the eigenstates is in the null space of the Liouvillian, if the qubits are positioned along the unidirectional axis such that the distance between them satisfies $k_q r_{2,1} = 2\pi n$, for $n$ integer. Therefore, this state $|\psi_s\rangle$ fulfills $\mathcal{L}[|\psi_s\rangle\langle\psi_s|] = 0$ and hence it is a dark state.

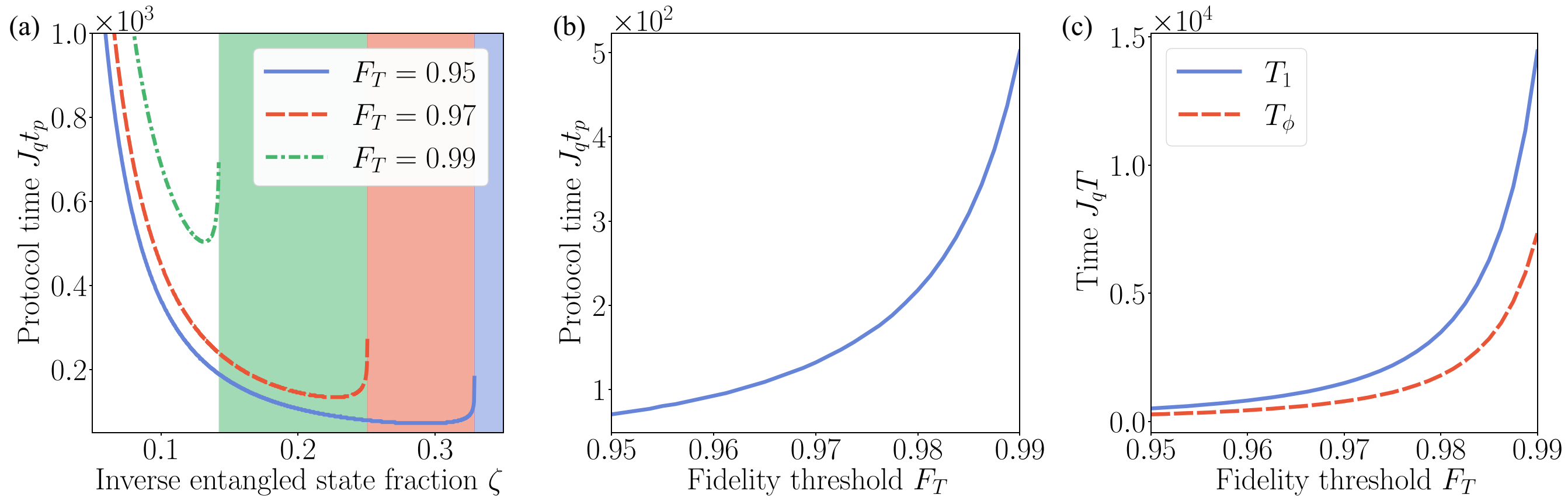

FIG. 3. (a) Protocol time $t_p$ normalized by the coupling strength $J_q$ as a function of $\zeta = J_q/(\sqrt{2}\Omega)$ for different fidelity thresholds $F_T$. A larger $\zeta$ corresponds to a shorter protocol time $t_p$, up to the regime where the threshold condition of Eq. (9) imposes an upper bound on $\zeta$, as given by Eq. (10) and indicated with a vertical line. A higher fidelity $F_T$ requires a longer time $t_p$ for a given $\zeta$. (b) Protocol time $t_p$ as a function of the fidelity threshold $F_T$. A higher fidelity comes at the cost of a longer time $t_p$. (c) Qubit lifetime $T_1$ and dephasing time $T_\phi$ normalized by the coupling strength $J_q$ as a function of the fidelity threshold $F_T$. A higher fidelity $F_T$ can be obtained at the expense of a longer required lifetime $T_1$ or dephasing time $T_\phi$.

Since this is the only state fulfilling the dark state condition, it is also the steady state. The existence of the steady state is guaranteed by the time independence of the Liouvillian [92,93]. For $J_q > 0$, the steady state takes the form

$$|\psi_s\rangle = \frac{1}{\sqrt{\zeta^2+1}}\left(i\zeta\,|00\rangle + \frac{1}{\sqrt{2}}(|10\rangle - |01\rangle)\right), \quad (8)$$

where $\zeta = J_q/(\sqrt{2}\Omega)$ [5]. For $\zeta \ll 1$, the steady state corresponds to the maximally entangled Bell state $|\psi_-\rangle = (|10\rangle - |01\rangle)/\sqrt{2}$. Thus, for sufficiently large driving with respect to the dissipative coupling, the steady state is maximally entangled (and pure), contrary to the bidirectional case (where the coupling constants in both directions are equal) where the steady state is not pure and not maximally entangled [7,8]. We study the case of directional coupling, where the coupling in one propagation direction is larger than the other, in Sec. III B. The parameter $\zeta$ controls the transient time $t_s$ required to reach the steady state, such that $t_s \to \infty$ for $\zeta \to 0$ (see Fig. 9 in Appendix D).

## III. PROTOCOL

We propose a protocol in which we let the state of two qubits separated by a distance $r_{2,1}$ with $k_q r_{2,1} = 2\pi n$ converge toward the entangled steady state given by Eq. (8). We take the initial state as $\hat{\rho}_{00} = |00\rangle\langle 00|$, since this is the steady state of the system without driving the qubits. After switching on the driving, we let this initial state evolve according to the Liouvillian in Eq. (4) for a given time $t$, giving the density matrix $\hat{\rho}(t)$. According to Eq. (8), the smaller $\zeta$, the higher the entangled state fraction of the steady state. We quantify the overlap between $\hat{\rho}(t)$ and the entangled state $|\psi_-\rangle$ by calculating the fidelity $F(t) = \sqrt{\langle\psi_-|\,\hat{\rho}(t)\,|\psi_-\rangle}$ [12]. For sufficient time passed and for appropriately chosen $\zeta$, the density matrix $\hat{\rho}(t)$ is such that the fidelity has exceeded a specific threshold $F_T$, i.e.,

$$F(t) = \sqrt{\langle\psi_-|\,e^{\mathcal{L}t}[\hat{\rho}_{00}]\,|\psi_-\rangle} \geqslant F_T. \quad (9)$$

We denote the minimal time needed to fulfill this condition as the protocol time $t_p$.

To limit the effects of the intrinsic qubit dissipation, the protocol time should be minimized. Therefore, we vary $\zeta$ to find the minimal protocol time $t_p$ to reach the fidelity threshold. One can see this optimization in Fig. 3(a) for $F_T = 0.95$, 0.97, and 0.99. A smaller $\zeta$ comes with a larger entangled state $|\psi_-\rangle$ fraction, but also with a larger protocol time $t_p$, following the trend of the transient time $t_s$. Thus, a larger $\zeta$ is preferred. However, if $\zeta$ is too large, the overlap of the steady state and the entangled state $|\psi_-\rangle$ becomes too small to surpass the fidelity threshold $F_T$. This upper bound of $\zeta$ as a function of the fidelity threshold can be determined by solving the relation $F_T = |\langle\psi_s|\psi_-\rangle|$ for the threshold value $\zeta_T$. This leads to

$$\zeta_T = \sqrt{\frac{1}{F_T^2} - 1}. \quad (10)$$

Thus, a larger fidelity threshold corresponds to a lower upper bound $\zeta_T$. Once $\zeta$ approaches the upper bound $\zeta_T$, the protocol time increases. Since the overlap of the steady state and entangled state decreases for increasing $\zeta$, the state of the system $\hat{\rho}(t)$ requires to have a higher overlap with the steady state, i.e., $\sqrt{\langle\psi_s|\hat{\rho}(t)|\psi_s\rangle}$, to surpass the fidelity threshold with the entangled state. A higher overlap of $\hat{\rho}(t)$ with the steady state typically corresponds to more time required (see also Fig. 9 in Appendix D). Thus, the protocol time increases for $\zeta$ close to $\zeta_T$. This results in a competition leading to a nonmonotonic behavior of the protocol time $t_p$ as a function of $\zeta$, together with an upper bound on $\zeta$, as shown in Fig. 3(a).

Varying $F_T$ and iterating over all $\zeta$, we compute the minimal protocol time $t_p$ for each $F_T$. In Fig. 3(b), we examine this relation between the fidelity and the protocol time. As expected, a higher fidelity comes at the cost of a higher protocol time. We investigate the effects of a variation in $\zeta$ on the protocol time and on the threshold fidelity in Appendix E.

As mentioned in Sec. II, the Liouvillian of Eq. (4) holds exactly in the zero-temperature limit. To benchmark a temperature threshold to work within this limit, we derived the effective qubit-qubit dynamics for nonzero temperature in Appendix B. In this case, terms arise on top of the Liouvillian presented in Eq. (4), which describe a thermal excitation of the qubits due to the coupling to the bath. The analysis in Appendix B shows that the protocol is robust in the low-temperature regime, where the finite-temperature qubit dynamics effectively reduce to the zero-temperature model. As temperature increases, thermally induced excitation processes degrade the fidelity reached at the protocol time $t_p$, so the state no longer matches the target entangled state as predicted by the zero-temperature model. Benchmarking the protocol for $F_T = 0.95$, 0.97, and 0.99 demonstrates that the zero-temperature description remains valid up to approximately $k_B\mathcal{T}/(\hbar\omega_q) \lesssim 0.2$. This therefore sets the operating temperature threshold; i.e., the protocol requires sufficiently cryogenic conditions to suppress thermal magnon occupation and preserve high-fidelity steady-state entanglement.

### A. Intrinsic qubit decoherence

Since spin qubits are intrinsically subject to dissipative processes, we investigate the effect of decay and dephasing of the spin qubits on the protocol. To include the former, we add the qubit decay $1/T_1 \sum_{i=1,2} \mathcal{D}[\hat{\sigma}_i^-]\hat{\rho}$, where $T_1$ is the qubit lifetime and

$$\mathcal{D}[\hat{A}]\hat{\rho} = \hat{A}\hat{\rho}\hat{A}^\dagger - \tfrac{1}{2}(\hat{A}^\dagger\hat{A}\hat{\rho} + \hat{\rho}\hat{A}^\dagger\hat{A}) \tag{11}$$

is the Lindblad dissipator. For a given fidelity threshold $F_T$, we increase $T_1$ and vary $\zeta$ until the fidelity threshold is surpassed, so that we obtain the minimal lifetime needed such that the protocol succeeds. The result is visualized in Fig. 3(c) with a blue solid line. We see that a higher threshold fidelity requires a higher lifetime. This is along the lines of the expectation, since a higher fidelity corresponds to a higher protocol time [see Fig. 3(a)] and hence the state has to be preserved for a longer time. In a similar fashion, we obtain a benchmark of the dephasing time $T_\phi$ by adding the qubit dephasing $1/T_\phi \sum_{i=1,2} \mathcal{D}[\hat{\sigma}_i^+\hat{\sigma}_i^-]\hat{\rho}$. Again, we vary $T_\phi$ until for any $\zeta$ the protocol converges. This results in the red dashed line in Fig. 3(c).

By comparing the lifetime $T_1$ and dephasing time $T_\phi$ in Fig. 3(c), we observe that the required lifetime is larger than the dephasing time for all fidelity thresholds $F_T$. We discuss the effect of dissipation on the protocol-time optimization given by Fig. 3(a) in Appendix G.

### B. Directional coupling

In general, we can distinguish three qubit-qubit coupling regimes: (1) unidirectional coupling, as treated above, where, e.g., $J_R$ finite and $J_L = 0$ (with $R$ and $L$ indicating the two possible directions), (2) directional coupling, where, e.g., $J_R > J_L$ but both finite, and (3) bidirectional coupling, where $J_R = J_L$. In this subsection, we investigate the intermediate regime of directional instead of unidirectional coupling of the qubits, which can happen when two magnon modes with wave numbers $k_{q,R}$ and $k_{q,L}$ fulfill the resonant condition $\omega_q = \omega_k$ (see Sec. II). We consider group velocities $v_{k_{q,R}} > 0$ and $v_{k_{q,L}} < 0$, corresponding to propagation toward the left and right, respectively, along the unidirectional axis. Due to the difference in sign, the coupling of the qubits to these magnon modes is not unidirectional. We assume that the dissipative coupling $J_{q,j} = 2\pi\mathcal{D}(k_{q,j})|g_{k_{q},j}|^2/|v_{k_{q},j}|$ due to the coupling to the magnon modes $k_{q,R}$ and $k_{q,L}$ is such that $J_{q,R} > J_{q,L}$. Thus, the coupling is directional.

The directional coupling is described by the Liouvillian $\mathcal{L}_{\text{di}} = \mathcal{L}_Q + \mathcal{L}_{\text{uni},1} + \mathcal{L}_{\text{uni},2}$, where $\mathcal{L}_Q[\hat{\rho}] = -(i/\hbar)[\hat{H}_Q, \hat{\rho}]$ and

$$\mathcal{L}_{\text{uni},j}[\hat{\rho}] = -\frac{i}{\hbar}(\hat{H}_{\text{eff},j}\hat{\rho} - \hat{\rho}\hat{H}_{\text{eff},j}^\dagger) + \hat{L}_j\hat{\rho}\hat{L}_j^\dagger, \tag{12}$$

with effective Hamiltonian $\hat{H}_{\text{eff},j} = \hat{H}_{\text{loc},j} + \hat{H}_{\text{uni},j}$ and Lindblad operator

$$\hat{L}_j = \sum_{i=1,2} \sqrt{J_{q,j}} e^{-ik_{q,j}r_i}\hat{\sigma}_i^-. \tag{13}$$

The first term of the effective Hamiltonian $\hat{H}_{\text{eff},j}$ is the local Hamiltonian

$$\hat{H}_{\text{loc},j} = -i\hbar \sum_{i=1,2} \frac{J_{q,j}}{2}\hat{\sigma}_i^+\hat{\sigma}_i^-. \tag{14}$$

The second term $\hat{H}_{\text{uni},j}$ describes the coupling of the qubits and depends on the propagation direction of the magnon mode, which for right-propagating magnons equals

$$\hat{H}_{\text{uni},R} = -i\hbar J_{q,R} e^{ik_{q,R}r_{2,1}}\hat{\sigma}_1^-\hat{\sigma}_2^+ \tag{15}$$

and for left-propagating magnons equals

$$\hat{H}_{\text{uni},L} = -i\hbar J_{q,L} e^{ik_{q,L}r_{1,2}}\hat{\sigma}_1^+\hat{\sigma}_2^-. \tag{16}$$

Now, we investigate the consequences of the directional coupling for the protocol. As we found previously for right-propagating magnons, the qubits should be positioned such that $k_{q,R}r_{2,1} = 2\pi n$, with $n$ integer, although this condition is not very stringent as shown in Appendix H. However, this condition may not be fulfilled for left-propagating magnons with wave number $k_{q,L}$. Therefore, we vary both $J_{q,L}$ and $k_{q,L}$ and examine their effect on the protocol time $t_{p,\text{di}}$. We do this in similar fashion to the unidirectional protocol case. We let the initial state $\hat{\rho}_{00}$ evolve in time according to the Liouvillian $\mathcal{L}_{\text{di}}$ until the fidelity threshold $F_T$ is surpassed. We show the results in Fig. 4, where we took $F_T = 0.95$. Here, we see that for $J_{q,L}/J_{q,R} \lesssim 10^{-4}$ the coupling $J_{q,L}$ does not prevent the convergence for all wave numbers $k_{q,L}$, where the directional protocol time $t_{p,\text{di}}$ is similar to the unidirectional protocol time $t_p$. For larger $J_{q,L}$, the convergence depends on the wave number $k_{q,L}$. If the phase $k_{q,L}r_{2,1}$ is not close to zero, or equivalently, a multiple of $2\pi$, then the fidelity threshold is not surpassed, which is indicated with white. Therefore, in the case of directional coupling, it is important to comply with this condition $k_{q,L}r_{2,1} = 2\pi m$, with $m$ integer. If this condition is met, the range of possible coupling strengths $J_{q,L}$ is vast as one can see in the inset of Fig. 4, where we take $k_{q,L}r_{2,1} = 0$. For $J_{q,L}/J_{q,R}$ approaching 1, we see that at some point there is no convergence, as we expect from the bidirectional ($J_{q,R}=J_{q,L}$) case [7,8].

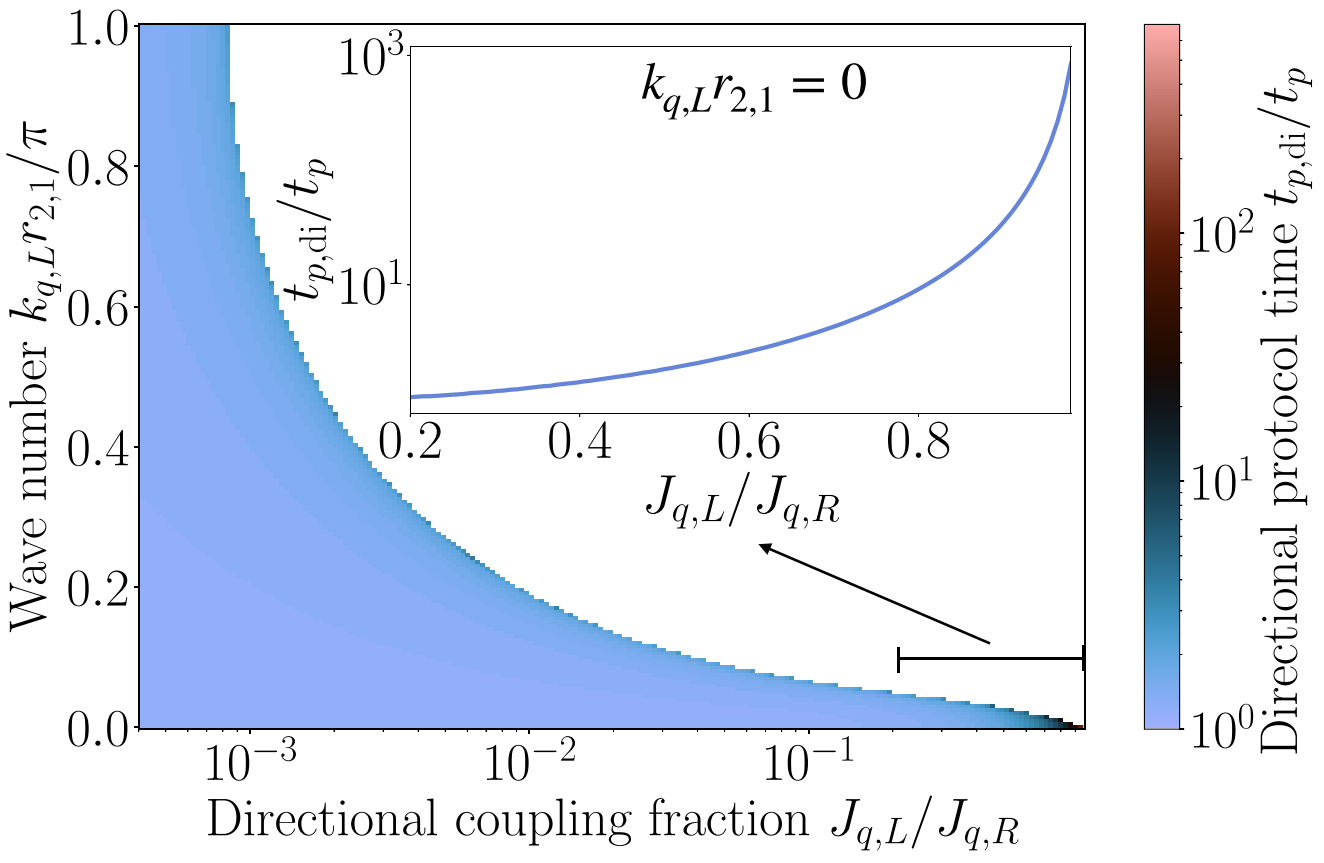

FIG. 4. Protocol time $t_{p,\mathrm{di}}$ with directional coupling normalized by the protocol time $t_p$ with unidirectional coupling as a function of the coupling to the left $J_{q,L}$ normalized by the coupling to the right $J_{q,R}$ and as a function of the phase $k_{q,L}r_{2,1}$ normalized by $\pi$. The white coloring corresponds to the case where the fidelity threshold is not surpassed. Sufficiently small coupling $J_{q,L}$ with respect to $J_{q,R}$ still leads to convergence for all wave numbers $k_{q,L}$, while for higher $J_{q,L}$ the convergence depends on $k_{q,L}$. The inset shows $t_{p,\mathrm{di}}$ as a function of the coupling $J_{-q}$ for $k_{q,L}r_{2,1}=0$. We use a maximal convergence time of $t_{p,\mathrm{di}}/t_p=1000$. The fidelity threshold is $F_T=0.95$, which corresponds to a protocol time $J_qt_p=73$ with $\zeta=0.27$ [see Fig. 3(a)].

## IV. IMPLEMENTATION WITH MAGNETIC SYSTEMS

In this section, we discuss in detail the implementation of the unidirectional coupling of single spins to chiral or nonreciprocal magnons, which we used in Sec. II. A schematic overview of possible physical mechanisms for unidirectional coupling in magnetic systems is given in Fig. 5. After discussing these mechanisms, we consider a specific hybrid system where this can be realized, consisting of NV centers coupled to the chiral and field displacement nonreciprocal surface waves of a YIG film.

### A. Mediation by chiral magnons

In this subsection, we discuss the interaction of a spin qubit with chiral magnons. Due to this chirality, the polarization of the magnons depends on their propagation direction. We show that the coupling constant of the qubits to the magnons depends on the magnon polarization, giving rise to a selection rule with respect to the magnon propagation direction, which leads to a unidirectional coupling.

The magnetic dipole moment $\hat{\boldsymbol{\mu}}_s=-\gamma_s\hat{\boldsymbol{S}}$ of the spin at position $\boldsymbol{r}_0$, where $\gamma_s$ is the modulus of the gyromagnetic ration, interacts with the magnetic field $\delta\hat{\boldsymbol{H}}(\boldsymbol{r})$ due to the magnons via

$$\hat{H}_{\mathrm{dipole}}=-\mu_0\hat{\boldsymbol{\mu}}_s\cdot\delta\hat{\boldsymbol{H}}(\boldsymbol{r}_0), \tag{17}$$

where $\mu_0$ is the vacuum magnetic permeability and

$$\delta\hat{\boldsymbol{H}}(\boldsymbol{r})=\sum_k(\delta\boldsymbol{H}_k(\boldsymbol{r})\hat{m}_k+\delta\boldsymbol{H}_k^*(\boldsymbol{r})\hat{m}_k^\dagger) \tag{18}$$

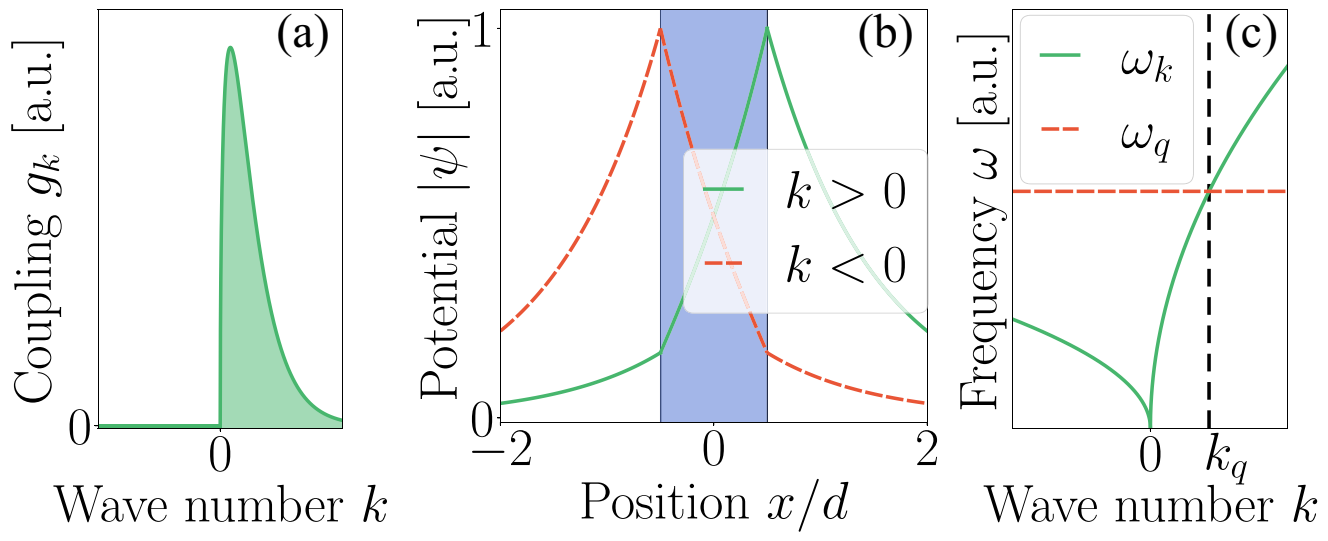

FIG. 5. Overview of the chirality and nonreciprocity leading to unidirectional spin-magnon coupling. (a) Due to chirality of the magnons and the polarization dependence of the spin-magnon coupling $g_k$, the coupling vanishes for one sign of the wave numbers $k$, but remains positive for the other, giving rise to unidirectional coupling. See Fig. 7(b) for an example. (b) Magnetic potential of a magnet exhibiting field displacement nonreciprocity (see Appendix I). The magnet is located between $x=-d/2$ and $d/2$, indicated in blue. Upon reversing the propagation direction $k\gtrless 0$, the magnetic potential shifts from one side of the magnet to the other. As a consequence, one propagation direction is more present on a specific surface than the other. (c) Sketch of the dispersion $\omega_k$ of a nonreciprocal magnon as a function of the wave number $k$. Due to the asymmetry in the dispersion, there is only one magnon wave number $k_q$ that is resonant with the qubit frequency $\omega_q$, and hence only one sign of $k$.

being the field fluctuation due to magnon mode $k$. We limit the description to the unidirectional axis of the magnet (see Sec. II).

The Cartesian spin components can be rewritten in terms of the spin ladder operators $\hat{S}_\pm=\hat{S}_x\pm i\hat{S}_y$ such that $\hat{\boldsymbol{S}}=(\hat{S}_+\boldsymbol{e}_-+\hat{S}_-\boldsymbol{e}_+)/\sqrt{2}+\hat{S}_z\boldsymbol{e}_z$, where $\boldsymbol{e}_\pm=(\boldsymbol{e}_x\pm i\boldsymbol{e}_y)/\sqrt{2}$. For spin 1/2, we express the spin raising and lowering operator in terms of the transition operators such that $\hat{S}_\pm^{S=1/2}=\hbar\hat{\sigma}^\pm$. We implement these relations and Eq. (18) in the Hamiltonian (17), on which we apply the RWA, giving

$$\hat{H}_{\mathrm{int}}^{S=1/2}=\frac{\hbar}{\sqrt{2}}\sum_k(g_{k,-}\hat{\sigma}^+\hat{m}_k+g_{k,-}^*\hat{m}_k^\dagger\hat{\sigma}^-), \tag{19}$$

where

$$g_{k,\alpha}=\gamma_s\mu_0\delta\boldsymbol{H}_k(\boldsymbol{r}_0)\cdot\boldsymbol{e}_\alpha. \tag{20}$$

The coupling constant $g_{k,-}$ vanishes if the magnon is left polarized ($\delta\boldsymbol{H}_k(\boldsymbol{r}_0)\propto\boldsymbol{e}_-$), contrary to right polarization $\boldsymbol{e}_+$ [94]. Therefore, for chiral magnons the spin-magnon coupling is unidirectional.

In the following, we will discuss an implementation with spin-1 qubits. For this case, three states $|0\rangle$, $|-\rangle$, and $|+\rangle$ are included in the description, which are eigenstates of the spin operator $\hat{S}_z$ with eigenvalues $m_s=0$, $-$, and $+$, respectively. The spin raising (lowering) operator $\hat{S}_+$ ($\hat{S}_-$) can be expressed in terms of the transition operators $\hat{\sigma}^{\alpha\beta}=|\alpha\rangle\langle\beta|$, such that $\hat{S}_\pm^{S=1}=\hbar\sqrt{2}(\hat{\sigma}^{\pm0}+\hat{\sigma}^{0\mp})$ [58]. Akin to the case of the spin 1/2, we plug these relations into the interaction Hamiltonian (17) and perform the RWA. We find

$$\hat{H}_{\mathrm{int}}^{S=1}=\hbar\sum_k(g_{k,+}\hat{\sigma}^{-0}+g_{k,-}\hat{\sigma}^{+0})\hat{m}_k+\mathrm{H.c.} \tag{21}$$

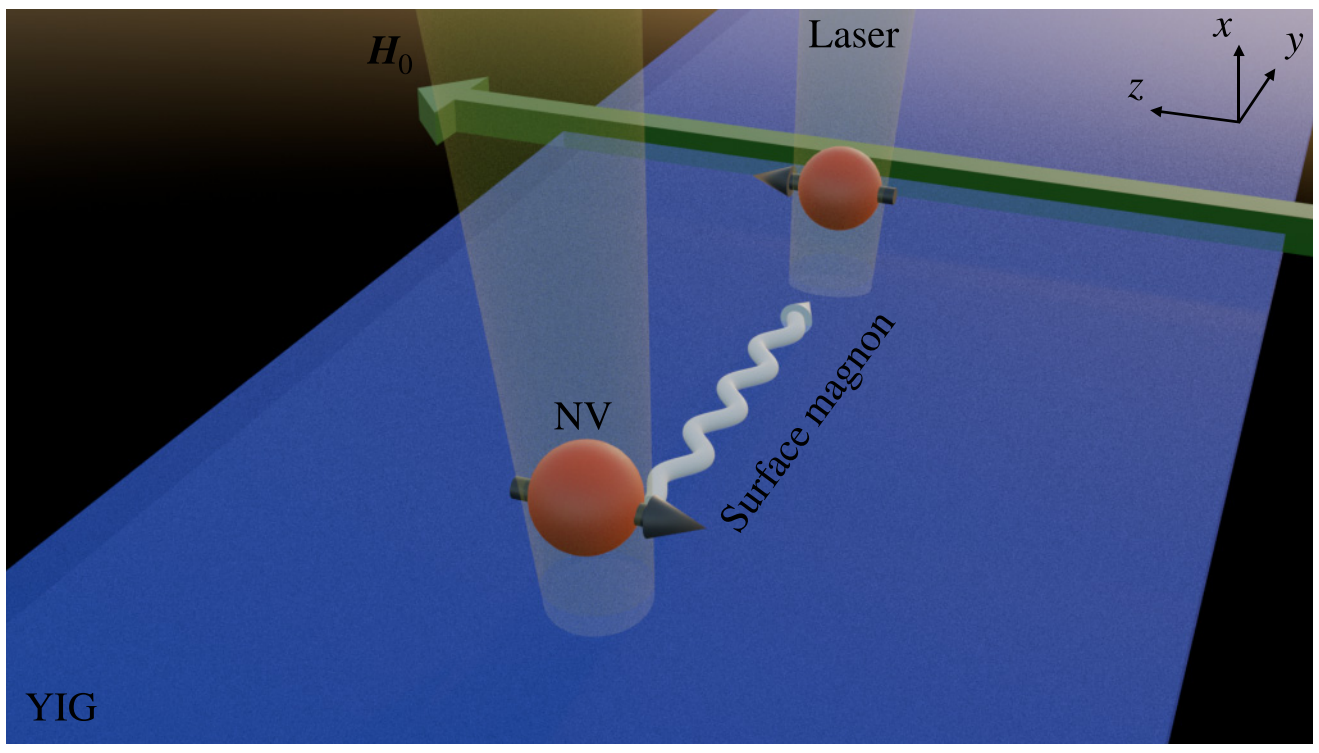

FIG. 6. Two NV centers driven at resonance by a laser are coupled unidirectionally through chiral and nonreciprocal surface magnons of a YIG slab. The propagation direction of these surface magnons is perpendicular to the external field $\boldsymbol{H}_0$.

If either $|-\rangle$ or $|+\rangle$ can be excluded from the dynamics of the coupled spins, then one can identify a qubit space that is spanned by the nonexcluded state together with the state $|0\rangle$. Since $g_{k,+}$ or $g_{k,-}$ are nonvanishing depending on the magnon polarization, this interaction also gives rise to a unidirectional coupling between magnons and a spin qubit with computational space $\{|0\rangle, |-\rangle\}$ or $\{|0\rangle, |+\rangle\}$.

Since the interaction Hamiltonians (19) for spin 1/2 and (21) for spin 1, after omitting the coupling of the $|0\rangle$ state to one of the triplet states $|+\rangle$ or $|-\rangle$, correspond to the Hamiltonian (3) of Sec. II and the unidirectionality condition is fulfilled, the effective qubit-qubit dynamics is described by the Liouvillian (4).

### B. Mediation by nonreciprocal magnons

In the previous subsection, we showed that the interaction between the spins and chiral magnons limits the coupling to only one propagation direction. Alternatively, the propagation of the magnons itself can be in certain cases unidirectional, and therefore the unidirectional coupling can be realized without resorting to chiral effects. We consider two scenarios for unidirectional propagation that can be found in magnets: field displacement and frequency nonreciprocity.

In the case of field displacement nonreciprocity, the mode profile of the magnons typically switches from one side of the magnetic structure to the other when inverting the propagation direction. The field displacement nonreciprocity was shown on a ferromagnetic slab [65,66], where the magnon modes are referred to as Damon-Eshbach modes. In Appendix I, we derive the magnetic potential $\psi$ of such a ferromagnetic slab, which is shown in Fig. 5(b). Here, we see that for one propagation direction of the magnons, the potential is localized on one side of the slab, while it switches to the other side upon reversing the propagation direction. Therefore, by positioning the spins on one side of the magnet, there is only one propagation direction of the magnons present to which the spins can couple (see also Fig. 6).

Frequency nonreciprocity refers to the asymmetry in the dispersion of the magnons depending on the propagation direction [see Fig. 5(c)]. There are different sources for this effect. The Dzyaloshinskii-Moriya interaction is one of those origins [67–69,75]. The dipolar interaction induces the nonreciprocity in curved ferromagnetic nanotubes [72,73] and for spin waves on ferromagnetic structures [65,66,95,96]. In multilayered structures, dipole-exchange interactions can give rise to asymmetries in the magnon dispersion [74,79,83,84]. Frequency nonreciprocity can also give rise to unidirectional magnon-spin coupling. By tuning the frequency of the spin so that it is in a regime where only one magnon mode is present, the resonance condition $\omega_k = \omega_q$ is only satisfied for this wave number, giving rise to a unidirectional dissipative coupling (see Sec. II).

If the dispersion in the case of field displacement nonreciprocity is symmetric, then the condition $k_{q,j} r_{2,1} = 2\pi n$ is automatically fulfilled for both wave numbers $k_{q,j}$ with $j = L, R$. Since we have shown the importance of complying with this condition in the case where the coupling is not purely unidirectional, this implies an advantage of field displacement nonreciprocity over frequency nonreciprocity.

### C. Example: steady-state entanglement mediated by chiral and nonreciprocal surface magnons

Here, we apply the findings of the previous sections on a ferromagnetic thin YIG film exhibiting both chiral and field displacement nonreciprocal properties coupled to NV centers. The configuration is shown in Fig. 6. NV centers are a prospective platform for quantum technologies due to their relatively long coherence times [89,97], and their integrability with other quantum systems, including photons and phonons. YIG is the magnetic material of choice in current state-of-the-art experiments, due to its record low dissipation values for magnons [87,98].

The ground-state manifold of an NV center consists of the triplet states $|0\rangle$, $|-\rangle$, and $|+\rangle$. The $|\pm\rangle$ states are energetically separated from the $|0\rangle$ state by the zero-field splitting $\hbar D_0$. The states $|\pm\rangle$ in turn can be split by an external magnetic field $H_0$, $\pm\omega_H$, with $\omega_H = \gamma_s \mu_0 H_0$. The Hamiltonian of the NV center reads

$$\hat{H}_{\mathrm{NV}} = \sum_{\eta=\pm} \hbar\omega_\eta \hat{\sigma}^{\eta\eta}, \tag{22}$$

with $\omega_\pm = D_0 \pm \omega_H$.

We consider a YIG film with thickness $d$ and area $L_y L_z \gg d^2$, where $L_y \gg L_z \gg d$. We apply an in-plane external field along the $z$ axis $\boldsymbol{H}_0 = H_0 \boldsymbol{e}_z$ (see Fig. 6). In Appendix I, we derive the frequency and the mode profiles of the surface magnons. The modes are chiral, with $\delta\boldsymbol{H}_k \propto \boldsymbol{e}_\mp$ for $k \gtrless 0$ on top of the film, and $\delta\boldsymbol{H}_k \propto \boldsymbol{e}_\pm$ for $k \gtrless 0$ on the bottom of the film, along the axis perpendicular to the in-plane external field, i.e., the $y$ axis. We neglect the higher-order magnon modes with $k_z > 0$ given that they are far detuned from the NV frequencies, $g_{n_z=1,k} \ll |\omega_{n_z=1,k} - \omega_q|$, where $n_z$ is defined with $k_z = 2\pi n_z / L_z$. This approximation enables us to focus on the surface magnons propagating along the $y$ axis with $k_z = 0$.

We consider two NV centers positioned on top of the YIG film along the unidirectional $y$ axis. Under the RWA, the interaction of the NV centers with the surface magnons is given by the interaction Hamiltonian of Eq. (21) [52,54–64]. The magnon frequency $\omega_k$ together with the frequencies $\omega_\pm$ of the qubit is displayed in Fig. 7(a). There are two magnon wave numbers $\pm k_-$ that are resonant with the NV frequency $\omega_- = \omega_{\pm k_-}$. This also holds for $\omega_+ = \omega_{\pm k_+}$, as one can see in Fig. 7(a). Note that the magnon wave number $k_+$ is significantly higher than $k_-$, as one can see in Fig. 7(a), due to the monotonic shape of the magnon dispersion and the fact that $\omega_+$ is larger than $\omega_-$.

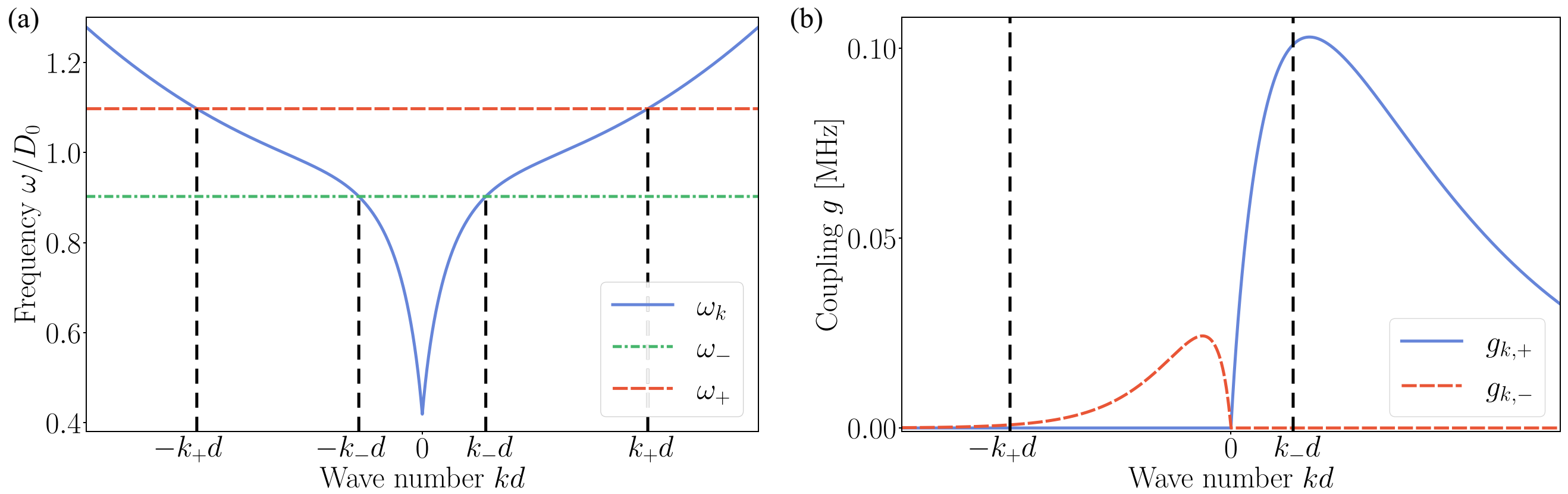

FIG. 7. (a) Frequencies $\omega_\pm$ of the NV states $|\pm\rangle$ and dispersion of the surface magnons $\omega_k$ along the unidirectional axis as a function of the magnon wave number $k$ normalized by the film thickness $d$. For each NV frequency $\omega_\pm$, there are two resonant wave numbers $\pm k_\pm$. (b) Coupling strengths $g_{k,\pm}$ as a function of the magnon wave number $k$ normalized by the film thickness $d$, which are limited to $k \gtrless 0$ due to the chirality of the magnons. Thus, the coupling is nonzero at only one of the resonant wave numbers $\pm k_\pm$ for each frequency $\omega_\pm$. The difference in amplitudes of $g_{k,\pm}$ is due to the field displacement nonreciprocity, which favors positive wave numbers (see Fig. 14). The parameters chosen to obtain these figures are in Table I.

Due to chirality, the NV-magnon coupling is unidirectional (see Sec. IV A), where the $|0\rangle \leftrightarrow |-\rangle$ transition of the NV is limited to coupling with $\boldsymbol{e}_-$ polarized magnons propagating to the right (with coupling constant $g_{k,+}$), while the transition $|0\rangle \leftrightarrow |+\rangle$ is limited to left-propagating magnons with polarization $\boldsymbol{e}_+$ ($g_{k,-}$). Due to the field displacement nonreciprocity, which favors right-propagating magnons, the coupling $g_{k,-}$ is suppressed with respect to $g_{-k,+}$ [see Fig. 7(b)]. This field displacement nonreciprocity together with the higher resonant wave number $k_+$ compared to $k_-$ gives rise to $g_{-k_+,-}/g_{k_-,+} \approx 10^{-2}$. Therefore, we can neglect $|0\rangle \leftrightarrow |+\rangle$ with respect to $|0\rangle \leftrightarrow |-\rangle$ to a good approximation [58]. Thus, we identify $|0\rangle$ and $|-\rangle$ as the computational basis states. From now on, we refer to the frequency $\omega_-$ as the qubit frequency $\omega_q$. Since the unidirectionality condition is fulfilled and the dynamics of the NV center can be truncated to two states, we obtain the Liouvillian given in Eq. (4).

Besides the instantaneous coupling conditions $|\delta_q|$, $J_q$, $\Omega \ll v_{k_q}/r_{2,1}$ [see Eq. (7)], the distance between the coupled qubits $r_{2,1}$ cannot exceed the coherence length of the magnons. For YIG films, the coherence length $l_m$ is limited by the lifetime of the magnons $\tau_m$. Using $l_m = v_{k_q}\tau_m$ and $\tau_m = 6\,\mu\text{s}$ [87] gives rise to $l_m = 3\,\text{mm}$. We check whether this coherence length is the limiting factor in the qubit distance $r_{2,1}$. Using the parameters given in Table I, we determine $J_q = 190\,\text{Hz}$. We note that there is no dependence of $J_q$ on $L_y$. We find $J_q/(v_{k_q}/l_m) = 10^{-3}$. Thus, the magnon coherence length is indeed the limiting factor. As a consequence, given that the qubit decoherence plays no role, the protocol can allow entangled spins with a distance exceeding microns, with the minimal qubit distance equaling $2\pi/k_q = 1.3\,\mu\text{m}$. This qubit distance is also such that the direct interaction between the NV centers is negligible [99]. For completeness, we note that $\tau_m/\tau_{\delta,\Omega} = 10^{-3}$ and $\tau_m \ll \tau_q$ [87,89,97].

To test the applicability of the protocol, we put the value of the dissipative coupling strength $J_q$ into context using the benchmark for the decoherence. Since the dephasing time is typically the critical factor for NV centers [86], we focus on this quantity in our discussion. A minimal fidelity of 0.95 requires a dephasing time $T_\phi = 280/J_q$ according to Fig. 3(c). For the determined coupling strength, this corresponds to $T_\phi = 1.5\,\text{s}$. Since the current dephasing times are on the order of seconds [89], this is the bottleneck of the protocol. We note that the protocol can be further combined with dynamical decoupling protocols [88,100].

Finally, we recall that to arrive at the Liouvillian in Eq. (4), we worked within the zero-temperature limit. In Appendix F, we verified that the findings of the protocol hold for $k_B\mathcal{T}/(\hbar\omega_q) \lesssim 0.2$. The frequencies of spin qubits are typically in the range 1–30 GHz [86,101], corresponding to a required temperature range 10–300 mK, which are routinely achievable in commercially available dilution fridges [24,37,38,40,102]. For a NV-center qubit specifically, the qubit frequency lies in the regime of the zero-field splitting $D_0 = 2\pi \times 2.877\,\text{GHz}$ [86], corresponding to a threshold temperature $\mathcal{T} = 28\,\text{mK}$ for the zero-temperature model to hold.

## V. CONCLUSIONS

We demonstrated that unidirectional coupling between magnons and spin qubits can be realized with magnets displaying field displacement or frequency nonreciprocity, chirality, or a combination of these. We showed that under the Born-Markov approximation the spin qubits are unidirectionally and dissipatively coupled. Exploiting this coupling, we proposed a protocol that evolves the product state $|00\rangle$ of two qubits into a maximally entangled Bell state. Since this state is the steady state of the system, any initial qubit state converges

to this entangled state, given that the distance $r$ between the two qubits satisfies $k_q r = 2\pi n$, where $k_q$ is the resonant magnon wave number and $n$ is an integer. We showed that the entanglement protocol remains effective beyond the perfectly unidirectional limit, provided the coupling is still strongly directional and the phase condition set by the qubit separation is satisfied. In the directional-coupling regime, a small backward coupling $J_{q,L}$ has little effect on the performance, but as $J_{q,L}/J_{q,R}$ increases, successful convergence becomes increasingly sensitive to the phase $k_{q,L}r_{2,1}$, and the protocol fails unless the phase stays close to an integer multiple of $2\pi$. Convergence also disappears as the system approaches the bidirectional limit. We discussed the application of the protocol on both chiral and nonreciprocal magnets.

We proposed and benchmarked the protocol for a hybrid system consisting of NV centers coupled via the surface magnons of a YIG film. Due to the high coherence length of magnons in YIG, steady-state entanglement can be achieved for qubits separated by a distance $r$ surpassing microns. We showed that the dephasing time $T_\phi$ of the NV centers is the limiting factor to achieve this entanglement in this hybrid system, requiring $T_\phi = 1.5$ s. The protocol requires to be performed at cryogenic temperatures to suppress the thermal occupation of the magnons, where we showed that a temperature $\mathcal{T} \lesssim 28$ mK suffices.

Stronger coupling between spin qubits and magnons helps to speed up the protocol, limiting the deteriorating effects of the intrinsic decoherence of the qubits. Recently, the strong coupling of an antiferromagnet (AFM) to electron spins has been demonstrated, where the electron spins are either doped directly into the AFM lattice [103] or hosted in a separate magnetic layer interfaced with the AFM [104]. Whereas the spin-magnon coupling of the rare-earth-doped AFM lacks nonreciprocity or chirality [103], such properties can be achieved by, e.g., magnetization grading [76] or by forming multilayered structures [80–84], providing an alternative promising platform for magnon-mediated steady-state entanglement. Alternatively, Ref. [104] demonstrated that the AFM magnons can be tuned *in situ* between linear and chiral modes, making this hybrid system suitable for the protocol and implying the possibility of combining it with bidirectional protocols.

Natural extensions to the protocol could involve harnessing the two-dimensional chiral coupling between emitters above a magnetic film. It has been shown [58] that for the right spatial configuration an emitter can couple chirally to emitters along not one but two well-defined directions, in analogy to a qubit placed at the intersection of two chiral waveguides. This could enable more complex physics and protocols such as the preparation of an entangled state between two qubits lying each on one of the effective "waveguides" by spontaneous decay of the original emitter. A second potential extension would be to explore the transition between coherent and dissipative coupling between emitters, either in purely magnonic or in hybrid magnonic-microwave platforms. Since the nature of the coupling depends on the wavelength of the magnons at the qubit energy, it should be possible to explore the transition between these two regimes by tuning the magnon dispersion relation. Recent works have shown that steady-state entanglement between spin qubits can be achieved by a symmetry-breaking interaction [105] and by postselection [106], or can be made more robust by bath engineering [107]. It would be interesting to investigate connections to these works by comparing with the unidirectional dynamics demonstrated in this work. Finally, the dissipative generation of singlet states of well-separated NV centers can find applications in quantum sensing [91,108], gradiometry, and distributed quantum computing applications [109]. As the total evolution of our open quantum system is governed by the non-Hermitian operator $\hat{H}_Q + \hat{H}_{\text{eff}}$, exceptional points (EPs) can be formed as shown in Fig. 8. The EP here is formed by the coalescence of two modes with varying singlet fraction [Eq. (C3)] and coalesce at $J_q/\Omega = 4$ (see Appendix C). The system's spectral response is maximally nonlinear in the vicinity of the EP, which could lead to higher-sensitivity quantum sensing applications [110,111]. Exploring these regimes is left for future work.

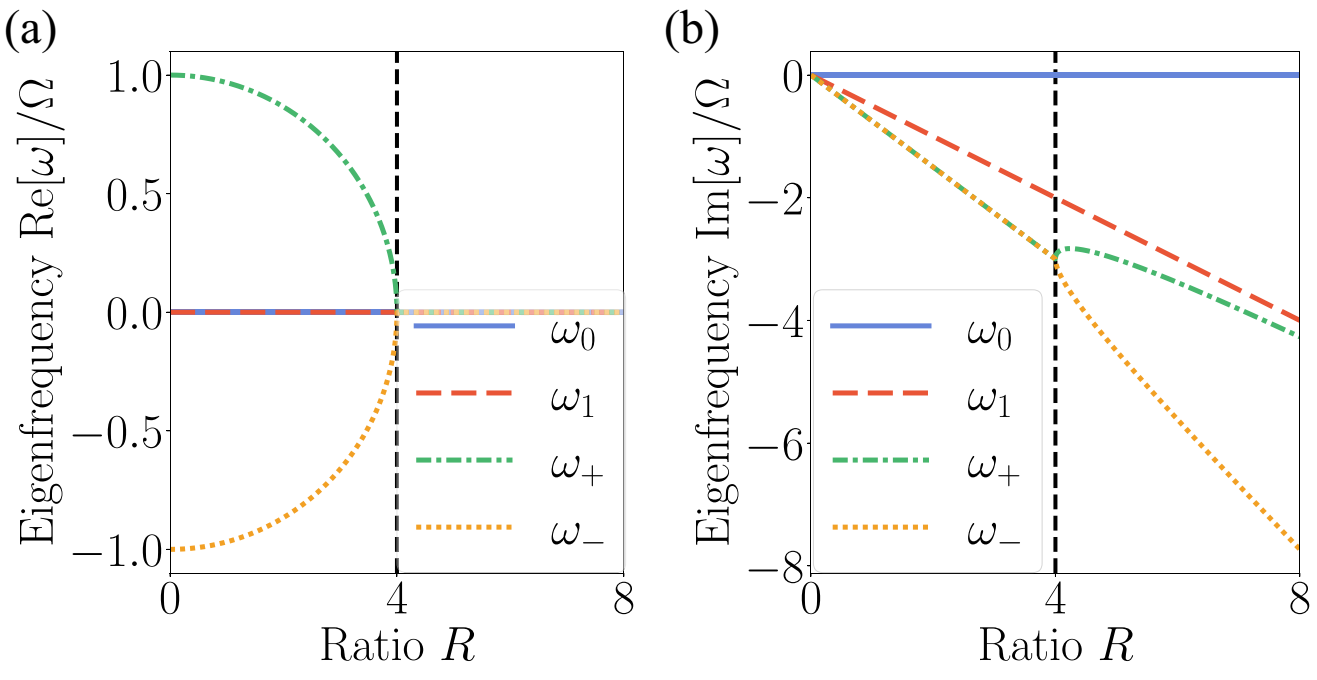

FIG. 8. (a) Real part and (b) imaginary part of the eigenfrequencies of $\hat{H}_Q + \hat{H}_{\text{eff}}$ as a function of $R = J_q/\Omega$ for two qubits and $\delta = 0$ and $k_q r_{2,1} = 2\pi n$. The exceptional point of the frequencies $\omega_\pm$ at $R = 4$ is indicated with a vertical dashed line.

## ACKNOWLEDGMENTS

M.D., M.C., D.B.R.D., and S.V.K. acknowledge financial support by the Federal Ministry of Research, Technology and Space (BMFTR) Project No. QECHQS (Grant No. 16KIS1590K). M.D., S.V.K. and M.C. acknowledge support from the EU-Project No. HORIZON-EIC-2021-PATHFINDEROPEN-01 PALANTIRI-101046630. C.G.-B. acknowledges the Austrian Science Fund FWF for the support with the Project No. PAT-1177623 "Nanophotonics-inspired quantum magnonics."

## DATA AVAILABILITY

The data that support the findings of this article are not publicly available. The data are available from the authors upon reasonable request.

## APPENDIX A: EFFECTIVE QUBIT-QUBIT DYNAMICS IN THE ZERO-TEMPERATURE LIMIT

In this Appendix, we use the Born-Markov approximation to trace out the magnon bath and derive the equation of motion of unidirectionally coupled spin qubits.

The Heisenberg equation of motion for the magnon annihilation operator $\hat{m}$ using the Hamiltonians (1) and (3) is given by

$$\frac{\mathrm{d}}{\mathrm{d}t}\hat{m}_{\mathbf{k}} = -i\delta_{\mathbf{k}}\hat{m}_{\mathbf{k}} - i\sum_j g^*_{j,\mathbf{k}}\hat{\sigma}^-_j. \tag{A1}$$

This equation is evaluated in the rotating frame of the drive, where $\delta_{\mathbf{k}} = \omega_{\mathbf{k}} - \omega_d$ is the detuning of the magnons. We solve the equation of motion (EOM) for the magnon annihilation operator formally to obtain

$$\hat{m}_{\mathbf{k}}(t) = \hat{m}_{\mathbf{k}}(0)e^{-i\delta_{\mathbf{k}}t} - i\sum_j g^*_{j,\mathbf{k}}\int_0^t \mathrm{d}\tilde{t}\hat{\sigma}^-_j(\tilde{t})e^{-i\delta_{\mathbf{k}}(t-\tilde{t})}. \tag{A2}$$

Plugging this relation into the EOM for a generic qubit operator $\hat{a}$, i.e., $\frac{\mathrm{d}}{\mathrm{d}t}\hat{a} = \frac{i}{\hbar}[\hat{H}_Q + \hat{H}_{\mathrm{int}}, \hat{a}]$, with $\hat{H}_Q$ and $\hat{H}_{\mathrm{int}}$ given in Eqs. (2) and (3), respectively, gives (all operators evaluated at time $t$)

$$\begin{aligned}\frac{\mathrm{d}}{\mathrm{d}t}\hat{a} = \frac{i}{\hbar}\Big([\hat{H}_Q, \hat{a}] + \sum_i[\hat{\sigma}^+_i, \hat{a}](\hat{\mathcal{N}}_i - i\hat{\mathcal{B}}_i)\\ + (\hat{\mathcal{N}}^\dagger_i + i\hat{\mathcal{B}}^\dagger_i)[\hat{\sigma}^-_i, \hat{a}]\Big),\end{aligned} \tag{A3}$$

where we identified a noise term

$$\hat{\mathcal{N}}_i = \sum_{\mathbf{k}}\hbar g_{i,\mathbf{k}}\hat{m}_{\mathbf{k}}(0)e^{-i\delta_{\mathbf{k}}t} \tag{A4}$$

and a backaction term

$$\hat{\mathcal{B}}_i = \sum_{j,\mathbf{k}}\hbar g_{i,\mathbf{k}}g^*_{j,\mathbf{k}}\int_0^t \mathrm{d}\tilde{t}\hat{\sigma}^-_j(\tilde{t})e^{-i\delta_{\mathbf{k}}(t-\tilde{t})}. \tag{A5}$$

We make the Markov approximation $\hat{\sigma}^-_j(\tilde{t}) \approx \hat{\sigma}^-_j(t)$, which is valid for $\tau_m \ll \tau_q, \tau_{\delta,\Omega}$, where $\tau_m$ and $\tau_q$ are the correlation times of the magnon bath and the spin qubit, respectively, and $\tau_{\delta,\Omega} = 2/\sqrt{\delta^2 + \Omega^2}$ is the time constant associated with the driving of the qubit. This gives rise to $\hat{\mathcal{B}}_i = \sum_j \hbar J_{i,j}\hat{\sigma}^-_j(t)$, where

$$J_{i,j} = \sum_{\mathbf{k}}\hbar g_{i,\mathbf{k}}g^*_{j,\mathbf{k}}\int_0^t \mathrm{d}\tilde{t}e^{-i\delta_{\mathbf{k}}(t-\tilde{t})}. \tag{A6}$$

We return to the EOM of Eq. (A3) and use $\mathrm{Tr}\{\frac{\mathrm{d}\hat{a}}{\mathrm{d}t}\hat{\rho}^H_{\mathrm{tot}}\} = \mathrm{Tr}\{\hat{a}\frac{\partial\hat{\rho}_{\mathrm{tot}}}{\partial t}\}$ to switch from the Heisenberg picture to the Schrödinger picture [2]. Here, $\hat{\rho}^H_{\mathrm{tot}}$ is the density operator for the complete system (so both qubits and magnons) in the Heisenberg picture. We employ the Born approximation; i.e., we consider that the qubit state is separable from the magnon state, which is assumed to be in the thermal state, i.e., $\hat{\rho}^H_{\mathrm{tot}} = \hat{\rho}^H_Q \otimes \hat{\rho}^H_{\mathrm{th}}$. Therefore, the trace over the magnonic part of the noise terms $\hat{\mathcal{N}}^{(\dagger)}_i$ will vanish. Given the Langevin equation in Eq. (A3), we can obtain a corresponding master equation by describing the magnon bath as a noise source. To identify the corresponding dynamics for the system state, it is necessary to first derive the appropriate stochastic equations, which follows the standard procedure in Ref. [112]. If we assume that the magnons constitute a zero-temperature bath, the dynamics of the qubits state $\hat{\rho}$ is given by

$$\frac{\partial\hat{\rho}}{\partial t} = -\frac{i}{\hbar}[\hat{H}_Q, \hat{\rho}] - \sum_{i,j} J_{i,j}[\hat{\sigma}^+_i, \hat{\sigma}^-_j\hat{\rho}] + J^*_{i,j}[\hat{\rho}\hat{\sigma}^+_j, \hat{\sigma}^-_i], \tag{A7}$$

where $\hat{\rho} = \mathrm{Tr}_M\{\hat{\rho}_{\mathrm{tot}}\}$. To arrive at this equation, we assumed that the interaction between qubits $i$ and $j$ is instantaneous. This requires $|\delta_q|$, $\Omega$, $|J_{i,j}| \ll v_m/r_{i,j}$, where $v_m$ is the group velocity of the magnons along the axis where the qubits are positioned. We note that in a physical implementation, the driving of the spin qubits would also give rise to a driving of the magnons. This would give rise to a nonvanishing average of the noise operator $\hat{\mathcal{N}}_i$. This in turn leads to a driving of the qubits [112], which can be absorbed into the driving of the qubits and hence does not change the resulting equation of motion (A7).

This master equation does not explicitly state the unidirectional coupling between the qubits. This comes into play when evaluating the coupling $J_{i,j}$. By substituting $t' = t - \tilde{t}$ and taking $t \to \infty$ [92], the time integral of $J_{i,j}$ in Eq. (A6) gives rise to [113]

$$\int_0^\infty \mathrm{d}t' e^{-i(\omega_{\mathbf{k}}-\omega_q)t'} = \pi\delta(\omega_{\mathbf{k}} - \omega_q) - i\mathcal{P}\left(\frac{1}{\omega_{\mathbf{k}} - \omega_q}\right). \tag{A8}$$

Here, we used $\omega_d = \omega_q$. We rewrite the coupling to $J_{i,j} = \gamma_{i,j}/2 + i\Omega_{i,j}$, where

$$\gamma_{i,j} = 2\pi\sum_{\mathbf{k}} g_{i,\mathbf{k}}g^*_{j,\mathbf{k}}\delta(\omega_{\mathbf{k}} - \omega_q) \tag{A9}$$

and

$$\Omega_{i,j} = -\sum_{\mathbf{k}} g_{i,\mathbf{k}}g^*_{j,\mathbf{k}}\mathcal{P}\left(\frac{1}{\omega_{\mathbf{k}} - \omega_q}\right). \tag{A10}$$

If one would plug this relation for $J_{i,j}$ into the EOM of Eq. (A7), one would identify $\gamma_{i,j}$ as a dissipative and $\Omega_{i,j}$ as a coherent coupling.

Now, we focus on the unidirectional axis. Using the resonant wave vectors $\pm k_q$, which are such that $\omega_{\pm k_q} = \omega_q$, the coupling $\gamma_{i,j}$ of Eq. (A9) equals

$$\begin{aligned}\gamma_{i,j} &= 2\pi\int \mathrm{d}k\mathcal{D}(k)g_{i,k}g^*_{j,k}\left(\frac{\delta(k-k_q)}{|v_{k_q}|} + \frac{\delta(k+k_q)}{|v_{-k_q}|}\right)\\ &= \frac{2\pi\mathcal{D}(k_q)|g_{k_q}|^2}{|v_{k_q}|}e^{ik_q r_{i,j}},\end{aligned} \tag{A11}$$

where we used the identity $\delta(f(x)) = \sum_i \frac{\delta(x-x_i)}{|f'(x_i)|}$, where $f(x_i) = 0$ and $f'(x_i) \neq 0$, and the fact that coupling is limited to one sign of $k$, such that $g_{i,-k_q} = 0$, in the case of chirality and field displacement nonreciprocity. In the case of frequency nonreciprocity, there is only one resonant wave vector $k_q$, such that the resulting $\gamma_{i,j}$ is identical. Also, we used $g_{i,k} = g_k\exp(ikr_i)$. Note that in reciprocal and symmetric ($\omega_k = \omega_{-k}$) systems, the dissipative coupling in Eq. (A9) would have another term involving $-k_q$. However, due to $g_{-k_q} = 0$ (chiral and field displacement nonreciprocity) or simply the lack of resonance for a wave number other than $k_q$ (frequency nonreciprocity), there is only the term giving rise to the resonance condition $\omega_{k_q} = \omega_q$. For the coherent coupling, one finds [114]

$$\Omega_{i,j} = -\frac{i\pi\mathcal{D}(k_q)\left|g_{k_q}\right|^2}{v_{k_q}} e^{ik_q r_{i,j}} \mathrm{sgn}(r_{i,j}). \tag{A12}$$

This yields for $v_{k_q} > 0$,

$$J_{i,j} = \frac{J_q}{2} e^{ik_q r_{i,j}}(1 + \mathrm{sgn}(r_{i,j})), \tag{A13}$$

where $J_q = 2\pi\mathcal{D}(k_q)|g_{k_q}|^2/|v_{k_q}|$. This results in

$$J_{i,i} = \frac{J_q}{2}, \quad J_{i>j} = J_q e^{ik_q r_{i,j}}, \quad J_{i<j} = 0 \tag{A14}$$

giving a coupling toward the right ($r_i > r_j$). Thus, the interplay of the dissipative coupling $\gamma_{i,j}$ of Eq. (A11) and coherent coupling $\Omega_{i,j}$ of Eq. (A12) gives rise to a unidirectional coupling $J_{i,j}$ [90]. For $v_{k_q} < 0$, plugging in $J_{i,j} = \gamma_{i,j}/2 + i\Omega_{i,j}$, one finds

$$J_{i,j} = \frac{J_q}{2} e^{ik_q r_{i,j}}(1 - \mathrm{sgn}(r_{i,j})), \tag{A15}$$

which results in a unidirectional coupling toward the left ($r_i < r_j$).

Rewriting the EOM (A7) yields

$$\frac{\partial\hat{\rho}}{\partial t} = -\frac{i}{\hbar}([\hat{H}_Q, \hat{\rho}] + \hat{H}_{\text{eff}}\hat{\rho} - \hat{\rho}\hat{H}^\dagger_{\text{eff}}) + \mathcal{J}[\hat{\rho}], \tag{A16}$$

with the non-Hermitian Hamiltonian

$$\hat{H}_{\text{eff}} = -i\hbar\sum_{i,j} J_{i,j}\hat{\sigma}_i^+\hat{\sigma}_j^- \tag{A17}$$

and the jump term

$$\mathcal{J}[\hat{\rho}] = \sum_{i,j} J_{i,j}\hat{\sigma}_j^-\hat{\rho}\hat{\sigma}_i^+ + J^*_{i,j}\hat{\sigma}_i^-\hat{\rho}\hat{\sigma}_j^+. \tag{A18}$$

With Eq. (A14), we find that the EOM of Eq. (A16) can be rewritten such that one obtains the Liouvillian of Eq. (4), where the Hamiltonian of Eq. (A17) gives rise to Eqs. (5) and (6), while the jump term of Eq. (A18) gives rise to the Lindblad term of Eq. (7).

## APPENDIX B: EFFECTIVE QUBIT-QUBIT DYNAMICS FOR FINITE TEMPERATURES

Here, we derive the effective qubit-qubit dynamics for finite temperatures. We refer to Ref. [92] for more details. Similar calculations where finite-temperature magnon baths are traced out can be found in Refs. [58,64]. While in the Appendix A we have described the system dynamics using the Heisenberg-Langevin approach, here we perform our calculations entirely in the Schrödinger picture and obtain directly a master equation for a finite-temperature magnon bath. This is more convenient, as the stochastic equations that one obtains from the Heisenberg-Langevin equations are less transparent and more complex to deal with. We note, however, that both approaches are equivalent [112].

The von Neumann equation for the density operator $\hat{\rho}^I_{\text{tot}}$ describing the total system in the interaction picture $\hat{\rho}^I_{\text{tot}}(t)$ is given by

$$\frac{\mathrm{d}}{\mathrm{d}t}\hat{\rho}^I_{\text{tot}}(t) = -\frac{i}{\hbar}\big[\hat{H}^I_{\text{int}}(t), \hat{\rho}^I_{\text{tot}}(t)\big], \tag{B1}$$

where $\hat{H}^I_{\text{int}}(t) = \hat{U}_I(t)\hat{H}_{\text{int}}\hat{U}_I^\dagger(t)$, with the unitary evolution operator $\hat{U}_I(t) = \exp(it(\hat{H}_M + \hat{H}_Q)/\hbar)$ and with the Hamiltonians $\hat{H}_M$, $\hat{H}_Q$, and $\hat{H}_{\text{int}}$ given in Eqs. (1)–(3), respectively. We employ the Born approximation, which states that the density matrix of the total system can be split such that $\hat{\rho}^I_{\text{tot}}(t) = \hat{\rho}^I(t)\otimes\hat{\rho}_{\text{th}}$, where $\hat{\rho}^I(t)$ describes the spin qubits and $\hat{\rho}_{\text{th}}$ is the thermal state describing the magnon bath. This approximation is valid provided that the coupling between the spin qubits and the magnons is small with respect to the qubit and magnon frequency, such that the qubits hardly influence the dynamics of the magnon bath. Integrating the von Neumann equation (B1) gives rise to

$$\hat{\rho}^I_{\text{tot}}(t) = \hat{\rho}^I_{\text{tot}}(0) - \frac{i}{\hbar}\int_0^t \mathrm{d}\tilde{t}\big[\hat{H}^I_{\text{int}}(\tilde{t}), \hat{\rho}^I_{\text{tot}}(\tilde{t})\big]. \tag{B2}$$

Plugging Eq. (B2) into Eq. (B1) yields

$$\begin{aligned}\frac{\mathrm{d}}{\mathrm{d}t}\hat{\rho}^I_{\text{tot}}(t) = &-\frac{i}{\hbar}\big[\hat{H}^I_{\text{int}}(t), \hat{\rho}^I(0)\otimes\hat{\rho}_{\text{th}}\big] \\ &-\frac{1}{\hbar^2}\int_0^t \mathrm{d}\tilde{t}\big[\hat{H}^I_{\text{int}}(t), \big[\hat{H}^I_{\text{int}}(\tilde{t}), \hat{\rho}^I_{\text{tot}}(\tilde{t})\big]\big].\end{aligned} \tag{B3}$$

Now, we trace over the magnonic degrees of freedom. Since the interaction $\hat{H}_{\text{int}}$ is linear in the magnon operator, one finds that the first term vanishes $\mathrm{Tr}_M\{[\hat{H}^I_{\text{int}}(t), \hat{\rho}^I(0)\otimes\hat{\rho}_{\text{th}}]\} = 0$. Then, we apply the Markov approximation to further simplify Eq. (B3). Here, one assumes that the bath decays much faster than the qubit, which is valid for $\tau_m \ll \tau_q$, where $\tau_m$ is the correlation time of the magnon bath and $\tau_q$ of the qubits. Also, we assume that the qubit driving is slow with respect to the correlation time $\tau_m$ of the magnon bath, $\tau_m \ll \tau_{\delta,\Omega} = 2/\sqrt{\delta^2+\Omega^2}$. Then, one can approximate $\hat{\rho}^I(\tilde{t}) \approx \hat{\rho}^I(t)$ and take the integral of Eq. (B3) from 0 to $\infty$ [92]. Making the substitution $\tilde{t} \to t - \tilde{t}$ gives rise to

$$\begin{aligned}\frac{\mathrm{d}}{\mathrm{d}t}\hat{\rho}^I(t) = &-\frac{1}{\hbar^2}\mathrm{Tr}_M \\ &\times\left\{\int_0^\infty \mathrm{d}\tilde{t}\big[\hat{H}^I_{\text{int}}(t), \big[\hat{H}^I_{\text{int}}(t-\tilde{t}), \hat{\rho}^I(t)\otimes\hat{\rho}_{\text{th}}\big]\big]\right\}.\end{aligned} \tag{B4}$$

The qubit drive is slow with respect to the correlation time of the bath ($\tau_m \ll \tau_{\delta,\Omega}$), such that the time propagation of the qubit is "frozen" during the interaction. Accordingly, the qubit operator $\hat{\sigma}_i^-$ remains unchanged in the interaction picture, while $\hat{m}_{\boldsymbol{k}}$ is transformed to $\hat{U}_I(t)\hat{m}_{\boldsymbol{k}}\hat{U}_I^\dagger(t) = \hat{m}_{\boldsymbol{k}}e^{-i\delta_{\boldsymbol{k}}t}$. Thus, one finds the following interaction Hamiltonian in the interaction picture:

$$\hat{H}^I_{\text{int}}(t) = \sum_{i,\boldsymbol{k}}\hbar(g_{\boldsymbol{k},i}\hat{\sigma}_i^+\hat{m}_{\boldsymbol{k}}e^{-i\delta_{\boldsymbol{k}}t} + \text{H.c.}). \tag{B5}$$

Inserting this Hamiltonian into Eq. (B4),

$$\begin{aligned}\frac{\mathrm{d}}{\mathrm{d}t}\hat{\rho}^I(t) = -\sum_{i,j}\big(&J^{\mathcal{T}}_{i,j}[\hat{\sigma}_i^+, \hat{\sigma}_j^-\hat{\rho}^I(t)] \\ &+ \big(J^{\mathcal{T}}_{i,j}\big)^*[\hat{\rho}^I(t)\hat{\sigma}_j^+, \hat{\sigma}_i^-] + \tilde{J}^{\mathcal{T}}_{i,j}[\hat{\rho}^I(t)\hat{\sigma}_j^-, \hat{\sigma}_i^+] \\ &+ \big(\tilde{J}^{\mathcal{T}}_{i,j}\big)^*[\hat{\sigma}_i^-, \hat{\sigma}_j^+\hat{\rho}^I(t)]\big),\end{aligned} \tag{B6}$$

where we defined

$$J_{i,j}^{\mathcal{T}} = \sum_{\mathbf{k}} \int_0^\infty d\tilde{t}(\bar{n}_{\mathbf{k}} + 1) g_{\mathbf{k},i} g_{\mathbf{k},j}^* e^{-i\delta_k \tilde{t}} \tag{B7}$$

and

$$\tilde{J}_{i,j}^{\mathcal{T}} = \sum_{\mathbf{k}} \int_0^\infty d\tilde{t} \bar{n}_{\mathbf{k}} g_{\mathbf{k},i} g_{\mathbf{k},j}^* e^{-i\delta_k \tilde{t}}. \tag{B8}$$

Also, we used $\mathrm{Tr}_M\{\hat{m}_{\mathbf{k}}\hat{m}_{\mathbf{k}'}\hat{\rho}_{\mathrm{th}}\} = \mathrm{Tr}_M\{\hat{m}_{\mathbf{k}}^\dagger\hat{m}_{\mathbf{k}'}^\dagger\hat{\rho}_{\mathrm{th}}\} = 0$, and $\mathrm{Tr}_M\{\hat{m}_{\mathbf{k}}\hat{m}_{\mathbf{k}'}^\dagger\hat{\rho}_{\mathrm{th}}\} = (\bar{n}_{\mathbf{k}} + 1)\delta_{\mathbf{k},\mathbf{k}'}$, as well as $\mathrm{Tr}_M\{\hat{m}_{\mathbf{k}'}^\dagger\hat{m}_{\mathbf{k}}\hat{\rho}_{\mathrm{th}}\} = \bar{n}_{\mathbf{k}}\delta_{\mathbf{k},\mathbf{k}'}$, with thermal occupation number $\bar{n}_{\mathbf{k}} = 1/(\exp(\hbar\omega_{\mathbf{k}}/k_B\mathcal{T}) - 1)$.

If one follows the same steps as provided in Appendix A, one finds

$$J_{i,j}^{\mathcal{T}} = (\bar{n}_{k_q} + 1)J_{i,j}, \quad \tilde{J}_{i,j}^{\mathcal{T}} = \bar{n}_{k_q}J_{i,j}, \tag{B9}$$

with $J_{i,j}$ given in Eq. (A13).

Now, we will rewrite the EOM (B6) in similar fashion to Appendix A. In the laboratory frame, one finds $\frac{d}{dt}\hat{\rho}(t) = \mathcal{L}_{\mathcal{T}}[\hat{\rho}]$, where

$$\mathcal{L}_{\mathcal{T}}[\hat{\rho}] = -\frac{i}{\hbar}([\hat{H}_Q, \hat{\rho}] + \hat{H}_{\mathrm{eff},\mathcal{T}}\hat{\rho} - \hat{\rho}\hat{H}_{\mathrm{eff},\mathcal{T}}^\dagger) + (\bar{n}_{k_q} + 1)\hat{L}\hat{\rho}\hat{L}^\dagger + \bar{n}_{k_q}\hat{L}^\dagger\hat{\rho}\hat{L}. \tag{B10}$$

Here, we defined

$$\hat{H}_{\mathrm{eff},\mathcal{T}} = (1 + \bar{n}_{k_q})\hat{H}_{\mathrm{eff}} + \bar{n}_{k_q}\hat{H}_{\mathrm{abs}}, \tag{B11}$$

with $\hat{H}_{\mathrm{eff}}$ as in Sec. II and in Eq. (A17) and with $\hat{H}_{\mathrm{abs}}$ describing thermal absorption

$$\hat{H}_{\mathrm{abs}} = -i\hbar\sum_i \frac{J_q}{2}\hat{\sigma}_i\hat{\sigma}_i^\dagger - i\hbar\sum_{i>j} J_q e^{-ik_q r_{i,j}}\hat{\sigma}_j^\dagger\hat{\sigma}_i. \tag{B12}$$

Thus, in the low-temperature limit $k_B\mathcal{T} \ll \hbar\omega_q$ and hence $\bar{n}_{k_q} \ll 1$, we retrieve the Liouvillian derived in Appendix A and presented in Eq. (4).

## APPENDIX C: EIGENSTATES OF THE HAMILTONIAN $\hat{H}_Q + \hat{H}_{\mathrm{eff}}$

To obtain the eigenstates of the Hamiltonian $\hat{H}_Q + \hat{H}_{\mathrm{eff}}$ for two resonantly driven qubits $\delta_q = 0$ and positioned such that $k_q r_{2,1} = 2\pi n$, we solve the equation $(\hat{H}_Q + \hat{H}_{\mathrm{eff}})|\psi_i\rangle = \hbar\omega_i|\psi_i\rangle$. Since the dimension of the Hilbert space is four, we obtain four eigenstates. The first eigenfrequency equals $\omega_0 = 0$ with eigenstate

$$|\psi_0\rangle \propto iR|00\rangle - |01\rangle + |10\rangle, \tag{C1}$$

with $R = J_q/\Omega > 0$. The second eigenfrequency equals $\omega_1 = -iJ_q/2$ with eigenstate

$$|\psi_1\rangle \propto |00\rangle - iR|01\rangle - |11\rangle, \tag{C2}$$

while the third and fourth eigenfrequencies equal $\omega_\pm = (-3iR/2 \pm \nu)\Omega/2$ with eigenstates

$$|\psi_\pm\rangle \propto \left(2 \pm i\nu R - \tfrac{1}{2}R^2\right)|00\rangle + \left(\tfrac{1}{2}iR(R^2 - 1) \pm \nu(R^2 + 1)\right)|01\rangle + \left(\tfrac{3}{2}iR \pm \nu\right)|10\rangle + (2 + R^2)|11\rangle. \tag{C3}$$

Here, $\nu = \sqrt{(4 - R)(4 + R)}/2$. For $R = 4$, one finds $\nu = 0$ and hence $\omega_+ = \omega_-$, while for all other $R$, one finds $\omega_+ \neq \omega_-$. Also, one finds that the eigenstates $|\psi_\pm\rangle$ coalesce for $R = 4$, since $\langle\psi_-|\psi_+\rangle = 1$. Therefore, $R = 4$ is an exceptional point.

In Fig. 8(a), we display the real and in Fig. 8(b) the imaginary parts of these eigenvalues. At the exceptional point $R = 4$, we see indeed that the eigenvalues $\omega_\pm$ are equal to each other.

Application of the Lindblad operator given in Eq. (7) on the eigenstate $|\psi_0\rangle$ of Eq. (C1) gives $\hat{L}|\psi_0\rangle = 0$. Since $|\psi_0\rangle$ is an eigenstate of $\hat{H}_Q + \hat{H}_{\mathrm{eff}}$ and is in the null space of the Lindblad operator, this state is a dark state of the Liouvillian of Eq. (4). Normalization of $|\psi_0\rangle$ gives rise to the state given in Eq. (8).

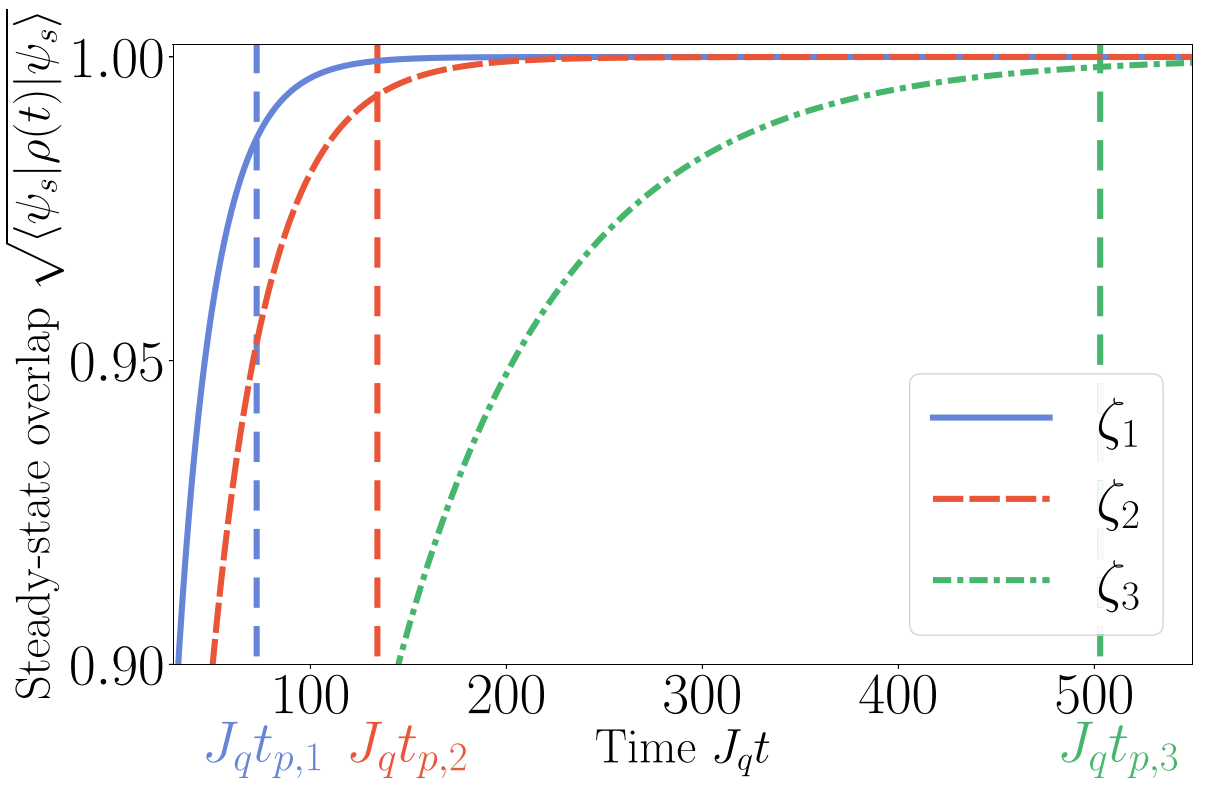

FIG. 9. Steady-state overlap $\sqrt{\langle\psi_s|\hat{\rho}(t)|\psi_s\rangle}$ of the system $\hat{\rho}(t)$ at the time $t$ normalized by the coupling constant $J_q$ for different $\zeta_1 > \zeta_2 > \zeta_3$. A higher steady-state overlap requires more time $t$. Also, a lower $\zeta$ corresponds to less time needed to reach the steady state and hence also to a lower transient time $t_s$. The vertical dashed lines correspond to the minimal protocol times $t_{p,i}$ found in Fig. 3(a) for a specific $\zeta_i$.

## APPENDIX D: TRANSIENT TIME VERSUS PROTOCOL TIME

We examine the time required for the system to reach the steady state, i.e., the transient time $t_s$. We quantify this by computing the overlap of the state of the system $\hat{\rho}(t)$ at time $t$ with the steady state $|\psi_s\rangle$, leading to $\sqrt{\langle\psi_s|\hat{\rho}(t)|\psi_s\rangle}$. The state $\hat{\rho}(t)$ is found by the master equation given by the Liouvillian in Eq. (4), using the initial state $\hat{\rho}_{00}$. We do this comparison for three cases, which are the minimal protocol times that we found in Fig. 3(a). These protocol times are $J_q t_{p,1} = 73$, $J_q t_{p,2} = 134$, and $J_q t_{p,3} = 503$, with the fidelity thresholds $F_{T,1} = 0.95$, $F_{T,2} = 0.97$, and $F_{T,3} = 0.99$. These values correspond to $\zeta_1 = 0.27$, $\zeta_2 = 0.22$, and $\zeta_3 = 0.13$, respectively. Note that $\zeta_1 > \zeta_2 > \zeta_3$. As one can see in Fig. 9, a lower $\zeta$ corresponds to more time $t$ required to reach a steady-state overlap ~1, giving rise to a higher transient time $t_s$.

We compare the transient time $t_s$ with the protocol time $t_p$ we find in Sec. III, i.e., the time needed for the system to be equal to the entangled state $|\psi_-\rangle$ within a specific fidelity threshold $F_T$. Also, note that at the dashed lines, which are the protocol times $t_{p,i}$ for a specific $\zeta_i$, that the system has not fully

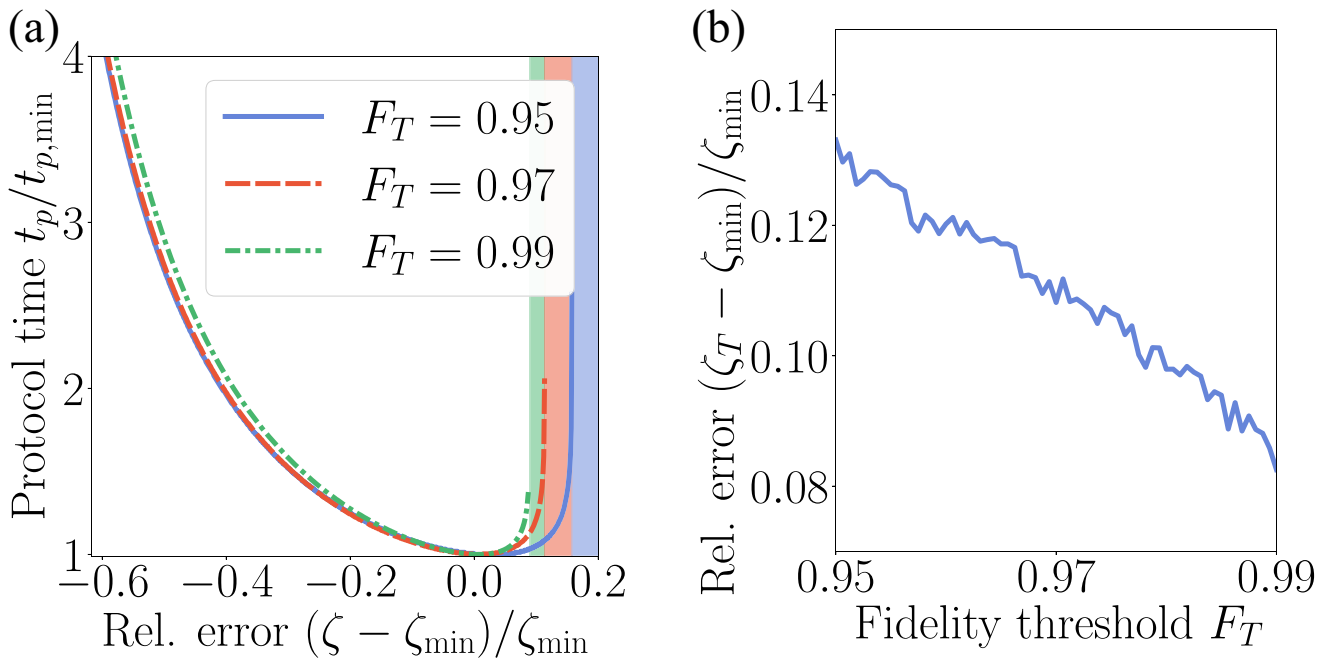

FIG. 10. (a) Protocol time $t_p$ normalized by the minimal time $t_{p,\min}$ as a function of the relative error of $\zeta$ with respect to the optimal $\zeta_{\min}$ that minimizes the protocol time. The threshold values $\zeta_T$ are indicated with a vertical line, which correspond to different fidelities $F_T$ [see Eq. (10)]. (b) Relative error of the threshold value $\zeta_T$ with respect to the optimal $\zeta_{\min}$ as a function of the fidelity threshold $F_T$.

converged yet toward the steady state, since the steady-state overlap is not equal to 1 yet.

## APPENDIX E: VARIATION OF THE PARAMETER $\zeta$

Here, we analyze the consequences of a variation in $\zeta = J_q/(\sqrt{2}\Omega)$. In the optimization of the protocol time $t_p$ versus $\zeta$ for a given fidelity threshold $F_T$, we find a $\zeta_{\min}$ that minimizes the protocol time, giving $t_{p,\min}$ [see Fig. 3(b)]. Here, we see that the protocol time is relatively flat around the minimum, indicating a wide range of $\zeta$ with a similar protocol time. To quantify this, we examine the variation in the protocol $t_p$ for a given (relative) error in $\zeta$ with respect to the optimal $\zeta_{\min}$, resulting in Fig. 10(a). Here, one sees that the allowed variation of $\zeta$ compared to the threshold value $\zeta_T$ depends on the set fidelity $F_T$. A higher fidelity $F_T$ comes at the cost of a lower allowed variation in $\zeta$ with respect to the threshold value $\zeta_T$. The threshold value $\zeta_T$ is relevant, because an error such that $\zeta$ surpasses the threshold $\zeta_T$ prohibits the protocol to converge. Therefore, we analyze this variation of $\zeta_T$ with respect to the optimal value $\zeta_{\min}$ as a function of the fidelity threshold in Fig. 10(b), setting an upper bound on the allowed error in $\zeta$. Since the relative error $(\zeta_T - \zeta_{\min})/\zeta_{\min}$ decreases for increasing fidelity $F_T$, we find once more that a higher fidelity requires a more precise configuration.

## APPENDIX F: VARIATION OF THE TEMPERATURE $\mathcal{T}$

As shown in Appendix B, for low temperatures $k_B\mathcal{T} \ll \hbar\omega_q$, the finite-temperature model of Appendix B corresponds to the zero-temperature model of Appendix A. Thus, for sufficiently low temperatures, one should retrieve the steady state given in Eq. (8). We vary the temperature $\mathcal{T}$ and benchmark the maximal temperature allowed such that the steady-state-entanglement protocol (see Sec. III) is valid. For this sake, we take the threshold fidelities $F_T = 0.95$, $F_T = 0.97$, and $F_T = 0.99$. In Fig. 3(a), we found that these fidelities correspond to the minimal protocol times $J_q t_p = 73$, 134, and 503, with corresponding $\zeta = 0.27$, 0.22, and 0.13, respectively. In analogy to the protocol, we let the input state $\hat{\rho}_{00} = |00\rangle\langle 00|$ propagate in time with the given $\zeta$ according to the Liouvillian $\mathcal{L}_\mathcal{T}$ and evaluate its fidelity with the entangled state $|\psi_-\rangle$ at the protocol time $t_p$. If the temperature is small enough, then the fidelity at the protocol time $t_p$ equals the threshold fidelity $F_T$, as prescribed by the zero-temperature model. As one can see in Fig. 11, for temperatures $k_B\mathcal{T}/(\hbar\omega_q) \lesssim 0.2$, the zero-temperature model holds.

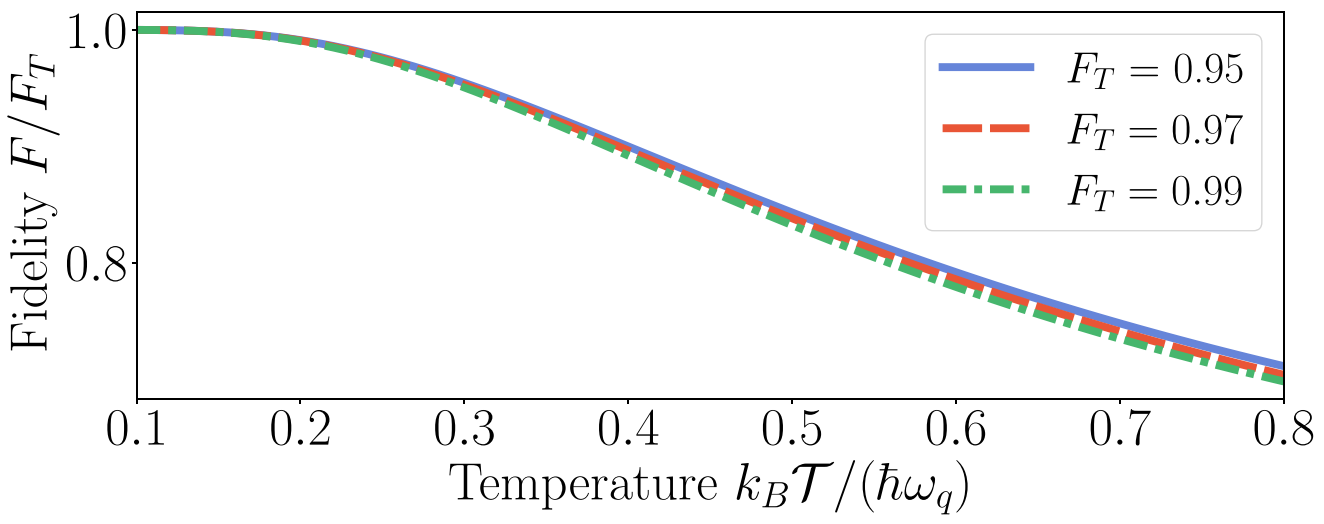

FIG. 11. Fidelity $F$ of the state propagated according to the Liouvillian $\mathcal{L}_\mathcal{T}$ at the protocol time $t_p$ normalized by the fidelity threshold $F_T$ expected in the zero-temperature model as a function of the temperature $\mathcal{T}$ normalized by $k_B/(\hbar\omega_q)$ for different fidelity thresholds $F_T$. The protocol time $t_p$ and $\zeta$ for each $F_T$ are given in the text of Appendix F.

## APPENDIX G: EFFECT OF QUBIT DECAY AND DEPHASING ON THE PROTOCOL-TIME OPTIMIZATION

Here, we discuss how adding dissipation changes the dependency of the protocol time $t_p$ on $\zeta$ for a fixed fidelity threshold $F_T = 0.95$, to be compared with the results without including dissipation as shown in Fig. 3(a). In Figs. 12(a) and 12(b), the blue line corresponds to the case where dissipation is negligible. Including a finite lifetime $T_1$ causes an increase of the time required for the steady state to reach the entangled state within the fidelity threshold. Also, the range of $\zeta$ for which there is convergence of the protocol is lower. A similar effect is found when including a dephasing time $T_\phi$. The addition of decay and dephasing changes the optimization, because dissipation is another "stationary attractor," leading to a different steady state. For example, for very short lifetimes of the qubits with respect to $1/J_q$, the steady state would be the $|00\rangle$ state. Thus, non-negligible dissipation changes the dependence of the protocol time on $\zeta$.

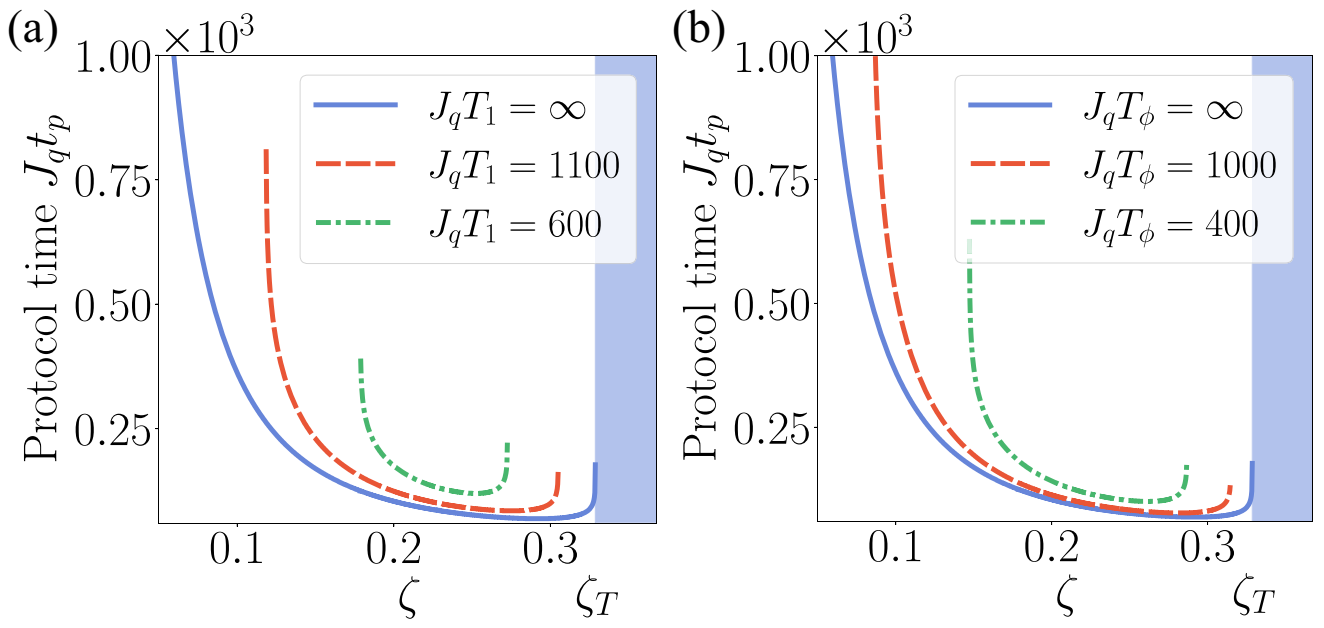

FIG. 12. Protocol time $t_p$ normalized by the coupling $J_q$ as a function of $\zeta$ for a fixed fidelity threshold $F_T = 0.95$ and for different lifetimes $T_1$ in panel (a) or dephasing times $T_\phi$ in panel (b).

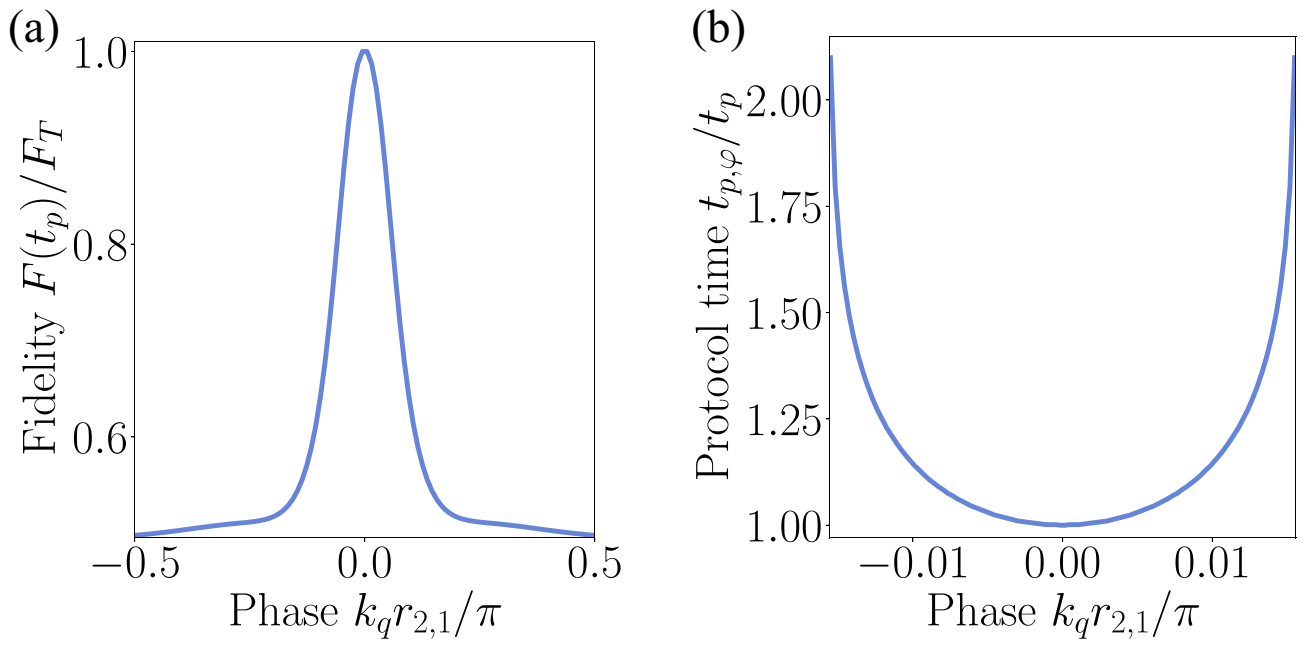

FIG. 13. (a) Fidelity $F(t_p)$ evaluated at the protocol time $t_p$, which is normalized by the fidelity threshold $F_T$ (see text) as a function of the phase $k_q r_{2,1}$. (b) Protocol time $t_{p,\varphi}$ normalized by the protocol time $t_p$ as a function of the phase $k_q r_{2,1}$.

## APPENDIX H: VARIATION OF THE QUBIT DISTANCE $r_{2,1}$

In this Appendix, we test the degree of variation in the qubit distance condition $k_q r_{2,1} = 2\pi n$ the protocol allows. As an example, we take $\zeta = 0.27$, $J_q t_p = 73$, which correspond to $F_T = 0.95$ (see Appendix D). First, we vary $k_q r_{2,1}$ and compute the fidelity $F(t_p)$, where $F(t)$ is the fidelity of the state $\hat{\rho}(t)$ of the system at time $t$ with the entangled state $|\psi_-\rangle$ [see Eq. (9)]. This results in Fig. 13(a). The fidelity $F(t_p)$ decreases more, The further the phase $k_q r_{2,1}$ varies from $2\pi n$, the more the fidelity $F(t_p)$ decreases. At $k_q r_{2,1} = \pi/2$, the fidelity reaches a minimum $F(t_p) = 1/2$, where the steady state of the system corresponds to the maximally mixed state $(\hat{\rho}_{00} + \hat{\rho}_{01} + \hat{\rho}_{10} + \hat{\rho}_{11})/4$ with $\hat{\rho}_{ij} = |ij\rangle\langle ij|$.

Since the time propagation of the system is stopped at a fixed time $t_p$ in the discussion above, where $k_q r_{2,1} = 0$ reaches the set fidelity threshold, the system does not reach this threshold for $k_q r_{2,1} \neq 0$. To investigate whether the system reaches the fidelity threshold at all, we fix the fidelity threshold to $F_T = 0.95$ as above and compute the time needed to surpass this threshold while varying the phase $k_q r_{2,1} \neq 0$, giving rise to $t_{p,\varphi}$ and Fig. 13(b). Here, we see that for $|k_q r_{2,1}| > 0.015 \times \pi$ the protocol does not converge, since the overlap of the steady state with the entangled state is insufficient. This imposes an upper bound on the admitted variation on $k_q r_{2,1}$.

## APPENDIX I: CHIRAL MAGNON SURFACE MODES

In this Appendix, we derive the magnon-spin coupling for a ferromagnetic film of thickness $d$ placed in a uniform, tangential magnetic field. We focus on the Damon-Eshbach configuration [65], in which spin waves propagate perpendicular to the field direction. In this case, the boundary condition for the normal component of the magnetic flux density $\boldsymbol{B}$ requires

$$(1+\chi_{\mathrm{d}})\left[\frac{\partial \psi^i}{\partial x} - i\chi_{\mathrm{a}}\frac{\partial \psi^i}{\partial y}\right]\bigg|_{x=\pm d/2} = \frac{\partial \psi^e}{\partial x}\bigg|_{x=\pm d/2} \tag{I1}$$

and the continuity of the tangential component of the magnetic field $\boldsymbol{H}$ implies

$$\psi^i|_{x=\pm d/2} = \psi^e|_{x=\pm d/2} + \text{const.} \tag{I2}$$

The additive constant in Eq. (I2) can be omitted. In Eq. (I1), $\chi_{\mathrm{d}}$ and $\chi_{\mathrm{a}}$ denote the diagonal and antidiagonal components of the Polder susceptibility tensor [115]. These components are given by

$$\chi_{\mathrm{d}} = \frac{\omega_{\mathrm{M}}\left(\omega_{\mathrm{H}} + \omega_{\mathrm{M}} k_y^2 D_{\mathrm{ex}}\right)}{\left(\omega_{\mathrm{H}} + \omega_{\mathrm{M}} k_y^2 D_{\mathrm{ex}}\right)^2 - \omega_k^2}, \tag{I3}$$

$$\chi_{\mathrm{a}} = \frac{\omega_k \omega_{\mathrm{M}}}{\left(\omega_{\mathrm{H}} + \omega_{\mathrm{M}} k_y^2 D_{\mathrm{ex}}\right)^2 - \omega_k^2}, \tag{I4}$$

where $\omega_{\mathrm{M}} = \gamma_s \mu_0 M_s$. The internal and external magnetic potentials, $\psi^i$ and $\psi^e$, satisfy Walker's and Laplace's equations, respectively:

$$(1+\chi_{\mathrm{d}})\left(\frac{\partial^2 \psi^i}{\partial x^2} + \frac{\partial^2 \psi^i}{\partial y^2}\right) + \frac{\partial^2 \psi^i}{\partial z^2} = 0, \tag{I5}$$

$$\frac{\partial^2 \psi^e}{\partial x^2} + \frac{\partial^2 \psi^e}{\partial y^2} + \frac{\partial^2 \psi^e}{\partial z^2} = 0. \tag{I6}$$

The corresponding solutions are assumed as

$$\psi^i = \left(A \sin\left(k_x^i x\right) + B\cos\left(k_x^i x\right)\right) e^{ik_y y}\cos(k_z z) \tag{I7}$$

in the ferromagnet $-d/2 \leqslant x \leqslant d/2$,

$$\psi^e = D e^{k_x^e x} e^{ik_y y}\cos(k_z z) \tag{I8}$$

below the ferromagnet $x < -d/2$, and

$$\psi^e = C e^{-k_x^e x} e^{ik_y y}\cos(k_z z) \tag{I9}$$

above the ferromagnet $x > d/2$. Substituting into Eqs. (I5) and (I6) yields the characteristic relations:

$$(1+\chi_{\mathrm{d}})\left[\left(k_x^i\right)^2 + k_y^2\right] + k_z^2 = 0, \tag{I10}$$

$$-\left(k_x^e\right)^2 + k_y^2 + k_z^2 = 0. \tag{I11}$$

We assume that the ferromagnet boundaries discretize the wave numbers along the $z$ axis as $k_z = 2\pi n_z/L_z$, where $n_z \in \mathbb{Z}$. Two types of magnon modes arise: volume and surface. The volume modes are characterized by real wave numbers, whereas the surface mode corresponds to an imaginary $k_x^i$ and real $k_y$, $k_z$ in the chosen geometry.

We focus on the surface mode with $k_z = 0$, assuming that the higher-order modes with $k_z > 0$ are far detuned from the qubit frequency,

$$g_{n_z=1,k_y} \ll |\omega_{n_z=1,k_y} - \omega_q|. \tag{I12}$$

This approximation allows us to restrict the analysis to magnons propagating along the $y$ axis with $k_z = 0$. From Eqs. (I10) and (I11), it then follows that

$$\left(k_x^i\right)^2 = -k_y^2, \tag{I13}$$

$$\left(k_x^e\right)^2 = k_y^2, \tag{I14}$$

which implies that $k_x^i$ is purely imaginary and continuous. We choose

$$k_x^i = ik_y, \tag{I15}$$

$$k_x^e = |k_y|. \tag{I16}$$

With this choice, the case $k_y > 0$ corresponds to a potential distribution that attains its maximum at the upper surface $x = d/2$ [see also Fig. 5(b)].

In addition, we find the dispersion equation that describes the magnon frequency

$$\omega_k^2 = \omega'_{\rm H}(\omega'_{\rm H} + \omega_{\rm M}) + \frac{\omega_{\rm M}^2}{2}\left(\frac{1}{2} - \frac{e^{-2|k_y|d}}{2}\right), \tag{I17}$$

where $\omega'_{\rm H} = \omega_{\rm H} + \omega_{\rm M} D_{\rm ex} k_y^2$.

Substituting the solutions into the boundary conditions [Eqs. (I1) and (I2)] yields

$$\begin{aligned}
&-iA\sinh\left(\frac{k_yd}{2}\right) + B\cosh\left(\frac{k_yd}{2}\right) - De^{-\frac{1}{2}|k_y|d} = 0,\\
&iA\sinh\left(\frac{k_yd}{2}\right) + B\cosh\left(\frac{k_yd}{2}\right) - Ce^{-\frac{1}{2}|k_y|d} = 0,\\
&[iAk_y(1+\chi_{\rm d}) + Bk_y\chi_{\rm a}]\cosh\left(\frac{k_yd}{2}\right) - [Bk_y(1+\chi_{\rm d}) + iAk_y\chi_{\rm a}]\sinh\left(\frac{k_yd}{2}\right) - D|k_y|e^{-\frac{1}{2}|k_y|d} = 0,\\
&\big[iAk_y(1+\chi_{\rm d}) + Bk_y\chi_{\rm a}\big]\cosh\left(\frac{k_yd}{2}\right) + \big[Bk_y(1+\chi_{\rm d}) + iAk_y\chi_{\rm a}\big]\sinh\left(\frac{k_yd}{2}\right) + C|k_y|e^{-\frac{1}{2}|k_y|d} = 0,
\end{aligned} \tag{I18}$$

where we employed Eqs. (I15) and (I16).

Solving Eq. (I18) yields

$$B = iA\frac{\tanh\left(\frac{k_yd}{2}\right)(|k_y| - \chi_{\rm a}k_y) + (1+\chi_{\rm d})k_y}{-\chi_{\rm a}k_y + |k_y| + (1+\chi_{\rm d})k_y\tanh\left(\frac{k_yd}{2}\right)}, \tag{I19}$$

$$C = iA\frac{(|k_y| - \chi_{\rm a}k_y)\sinh(k_yd) + (1+\chi_{\rm d})k_y\cosh(k_yd)}{(|k_y| - \chi_{\rm a}k_y)\cosh\left(\frac{k_yd}{2}\right) + (1+\chi_{\rm d})k_y\sinh\left(\frac{k_yd}{2}\right)}e^{\frac{1}{2}|k_y|d}. \tag{I20}$$

Next, we calculate the magnetization fluctuation using

$$\delta\boldsymbol{M} = \begin{pmatrix}\chi_{\rm d} & -i\chi_{\rm a} & 0\\ i\chi_{\rm a} & \chi_{\rm d} & 0\\ 0 & 0 & 0\end{pmatrix}\nabla\psi, \tag{I21}$$

yielding

$$\delta M_x = \big[(A\chi_{\rm a}k_y - B\chi_{\rm d}k_x^i)\sin(k_x^ix) + (A\chi_{\rm d}k_x^i + B\chi_{\rm a}k_y)\cos(k_x^ix)\big]e^{ik_yy}\cos(k_zz), \tag{I22}$$

$$\delta M_y = i\big[(A\chi_{\rm d}k_y - B\chi_{\rm a}k_x^i)\sin(k_x^ix) + (A\chi_{\rm a}k_x^i + B\chi_{\rm d}k_y)\cos(k_x^ix)\big]e^{ik_yy}\cos(k_zz), \tag{I23}$$

and $\delta M_z = 0$. These expressions are substituted into the magnon normalization condition

$$\int {\rm d}V\,\boldsymbol{M}_{\rm s}\cdot(\delta\boldsymbol{M}^*\times\delta\boldsymbol{M}) = i\gamma\hbar M_{\rm s}^2, \tag{I24}$$

leading to

$$2L_yL_zk_y(B\chi_{\rm d} + iA\chi_{\rm a})(A\chi_{\rm d} - iB\chi_{\rm a})\sinh(k_yd) + i\gamma\hbar M_{\rm s} = 0, \tag{I25}$$

where $L_yL_z$ denotes the cross-sectional area of the magnet. Substituting Eq. (I19) gives

$$A = -\,{\rm sgn}(k_y)\sqrt{\frac{\gamma\hbar M_{\rm s}}{2L_yL_zk_y\mathfrak{a}_1\mathfrak{a}_2\sinh(k_yd)}}\left\{{\rm sgn}(k_y)\left[\chi_{\rm a} + (1+\chi_{\rm d})\tanh\left(\frac{k_yd}{2}\right)\right] + 1\right\}, \tag{I26}$$

with

$$\mathfrak{a}_1 = \big[\chi_{\rm a} - (\chi_{\rm d}^2 + \chi_{\rm d} - \chi_{\rm a}^2)\,{\rm sgn}(k_y)\big]\tanh\left(\frac{k_yd}{2}\right) + \chi_{\rm a}\,{\rm sgn}(k_y) - \chi_{\rm d},$$

$$\mathfrak{a}_2 = [\chi_{\rm a}\,{\rm sgn}(k_y) - \chi_{\rm d}]\tanh\left(\frac{k_yd}{2}\right) - (\chi_{\rm d}^2 + \chi_{\rm d} - \chi_{\rm a}^2)\,{\rm sgn}(k_y) + \chi_{\rm a}.$$

The associated magnetic-field fluctuations at the upper surface $x > d/2$ are given by $\delta\boldsymbol{H} = \nabla\psi$, yielding

$$\delta\boldsymbol{H}_k(\boldsymbol{r}) = \delta H_k(\boldsymbol{r})\times\begin{cases}\boldsymbol{e}_-, & k_y > 0,\\ \boldsymbol{e}_+, & k_y < 0.\end{cases} \tag{I27}$$

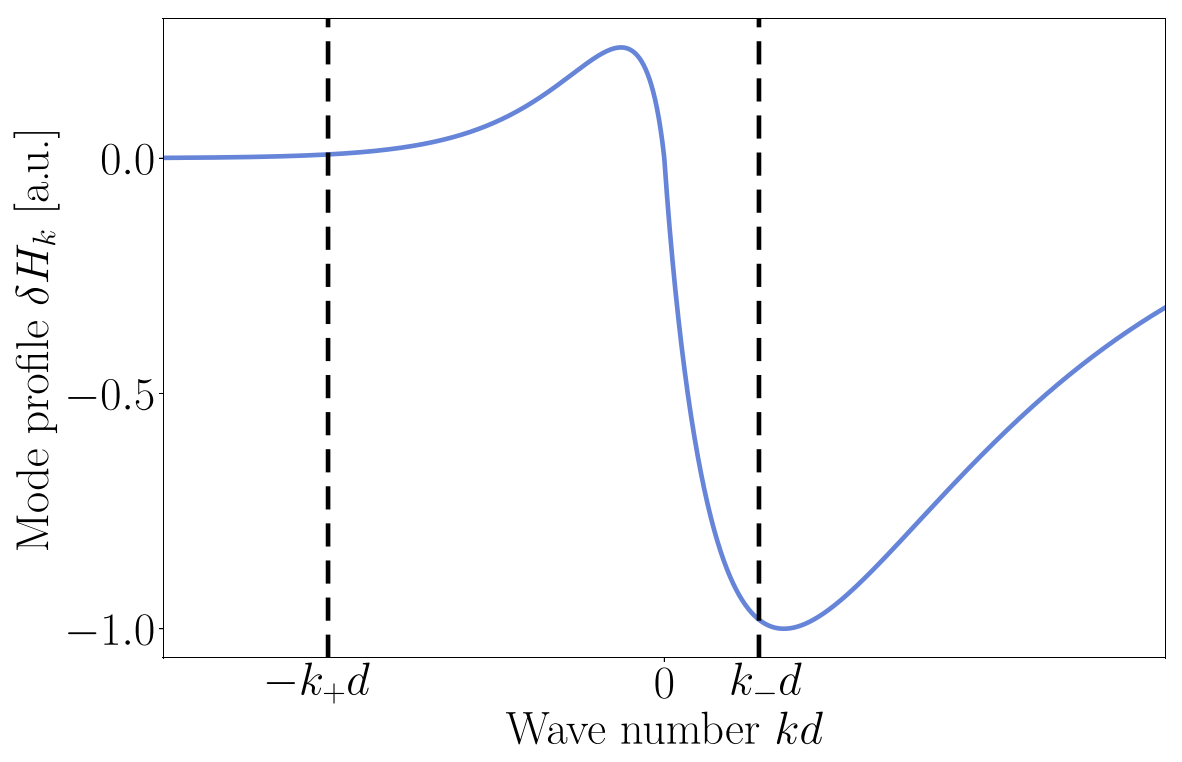

FIG. 14. Normalized mode profile $\delta H_k$ as a function of the wave number $k$ normalized by the film thickness $d$ evaluated at $x = d_{\rm NV} + d/2$ and $y = z = 0$. Due to the field displacement nonreciprocity, the mode profile has a higher amplitude for positive $k$ than for negative $k$. The resonant wave numbers $-k_+$ and $k_-$ from Fig. 7 are also shown. The parameters chosen to obtain this figure are in Table I.

Here, we have $\delta H_k(\boldsymbol{r}) = \sqrt{2}Ck_y e^{-x|k_y|+ik_y y}$ (see Fig. 14). Note that we consider the single propagation direction. Thus, the subscript $y$ in the wave number is omitted hereafter, i.e., $k_y = k$.

The magnon-qubit coupling $g_k$ along the unidirectional axis is determined by plugging the mode profile of Eq. (I27) evaluated at the qubit position into Eq. (20). We use the dispersion in Eq. (I17) to determine the group velocity $v_k$ and use $\mathcal{D}(k) = L_y/(2\pi)$ to determine the dissipative coupling $J_q = 2\pi\mathcal{D}(k_q)|g_{k_q}|^2/|v_{k_q}|$, with the parameters listed in Table I.

TABLE I. Summary of the parameters used to estimate the NV-magnon coupling $g_{k,\pm}$ and the dissipative coupling $J_q$.

| Parameter | Value |
|---|---|
| NV zero-field splitting | $D_0 = 2\pi \times 2.877$ GHz |
| Gyromagnetic ratio | $\gamma_s = 1.76 \times 10^{11}\ \mathrm{T^{-1}\,s^{-1}}$ |
| YIG film thickness | $d = 200$ nm |
| YIG film length | $L_y = 10$ µm |
| YIG film width | $L_z = 1$ µm |
| YIG exchange stiffness | $D_{\rm ex} = 3.086 \times 10^{-16}\ \mathrm{m^2}$ |
| YIG saturation magnetization | $M_s = 1.39 \times 10^5\ \mathrm{A\,m^{-1}}$ |
| YIG surface to NV distance | $d_{\rm NV} = d/2$ |
| External field amplitude | $\mu_0 H_0 = 10^{-2}$ T |